\documentclass[10pt,conference,letterpaper]{IEEEtran}
\pdfoutput=1
\usepackage{times,amsmath,epsfig}
\usepackage{amssymb}
\usepackage{url}
\usepackage{graphicx}
\usepackage{subfigure}
\usepackage{enumerate}
\usepackage{multirow} 
\usepackage{slashbox} 
\usepackage{algorithmic}
\usepackage{algorithm2e}

\newtheorem{definition}{Definition}
\newtheorem{example}{Example}

\newcommand{\ka}{$k\mbox{-}$anonymity}
\newcommand{\dld}{distinct $l\mbox{-}$diversity}
\newcommand{\tc}{$t\mbox{-}$closeness}
\newcommand{\bt}{$(B,t)\mbox{-}$privacy}
\newcommand{\js}{\emph{JS} divergence}
\newcommand{\defense}{{JS-reduce}}
\newcommand{\Hi}{\mathcal{H}}

\newcommand{\bksv}{BK^{sv}}
\newcommand{\bkseq}{BK^{seq}}
\newcommand{\spmbkseq}{{\textrm{\emph{SPM-}}}\bkseq}
\newcommand{\dkbkseq}{{\textrm{\emph{DK-}}}\bkseq}
\newcommand{\iebkseq}{{\textrm{\emph{IE-}}}\bkseq}

\newcommand{\pksv}{PK^{sv}}
\newcommand{\rbksv}{RBK^{sv}}
\newcommand{\QIG}{QI-group}
\newcommand{\qi}{QI}

\newcommand{\squishlist}{
   \begin{list}{$\circ$}
    { \setlength{\itemsep}{0pt}      \setlength{\parsep}{3pt}
      \setlength{\topsep}{3pt}       \setlength{\partopsep}{0pt}
      \setlength{\leftmargin}{1.5em} \setlength{\labelwidth}{1em}
      \setlength{\labelsep}{0.5em} } }
\newcommand{\squishlisttwo}{
   \begin{list}{$\bullet$}
    { \setlength{\itemsep}{0pt}    \setlength{\parsep}{0pt}
      \setlength{\topsep}{0pt}     \setlength{\partopsep}{0pt}
      \setlength{\leftmargin}{2em} \setlength{\labelwidth}{1.5em}
      \setlength{\labelsep}{0.5em} } }
\newcommand{\squishend}{
    \end{list}  }

 \graphicspath{{figure/}}

\title{Preserving Privacy in Sequential Data Release against Background Knowledge Attacks}
\author{%
{Daniele Riboni, Linda Pareschi, Claudio Bettini}%
\vspace{1.6mm}\\
\fontsize{10}{10}\selectfont\itshape
Universit\`a degli Studi di Milano, D.I.Co.\\
via Comelico 39, I-20135 Milano, Italy\\
\fontsize{9}{9}\selectfont\ttfamily\upshape
\{riboni,pareschi,bettini\}@dico.unimi.it%
}
\begin{document}
\maketitle
\begin{abstract} 
  A large amount of transaction data containing associations between
  individuals and sensitive information flows everyday into data
  stores. Examples include web queries, credit card transactions,
  medical exam records, transit database records. The serial release
  of these data to partner institutions or data analysis centers is a
  common situation.  In this paper we show that, in most domains,
  correlations among sensitive values associated to the same
  individuals in different releases can be easily mined, and used to
  violate users' privacy by adversaries observing multiple data
  releases.  We provide a formal model for privacy attacks based on
  this sequential background knowledge, as well as on background
  knowledge on the probability distribution of sensitive values over
  different individuals.  We show how sequential background knowledge
  can be actually obtained by an adversary, and used to identify with
  high confidence the sensitive values associated with an individual.
  A defense algorithm based on Jensen-Shannon divergence is proposed,
  and extensive experiments show the superiority of the proposed
  technique with respect to other applicable solutions.  To the best
  of our knowledge, this is the first work that systematically
  investigates the role of sequential background knowledge in serial
  release of transaction data.
\end{abstract}

\section{Introduction}
\label{sec:introduction}
Large amounts of transaction data related to individuals are
continuously acquired, and stored in the repositories of industry and
government institutions.  Examples include online service requests,
web queries, credit card transactions, transit database records,
medical exam records.  These institutions often need to repeatedly
release new or updated portions of their data to other partner
institutions
for different purposes, 
including distributed processing, participation in inter-organizational workflows, 
and data analysis. 
The medical domain is an interesting example: 
many countries have recently established centralized data stores that
exchange patients' data with medical institutions; 
new records are periodically released to data analysis centers in
non-aggregated form.

A very challenging issue in this scenario is the protection of
users' privacy, %
considering that potential adversaries have access to multiple serial
releases and can easily acquire background knowledge related to the
specific domain. This knowledge includes the 
fact that 
certain sequences of values in subsequent releases are more likely to be observed than other sequences. 
For example, it is pretty straightforward to extract from the medical
literature or from a public dataset that a sequence of medical exam results within a certain
time frame has higher probability to be observed than another
sequence. 
\begin{table*}[t!]
\label{tbl:rel}
\caption{Original and generalized transaction data at the first and second release 
(first and second week, respectively)}
\begin{footnotesize}
\begin{center}
\subtable[Original transaction data at time $\tau_1$]{\label{tbl:or1}
\begin{tabular}{|r|r|r|r|r|}
\hline
{\bf Name} &  {\bf Age} & {\bf Gender} &  {\bf Zip} & {\it {\bf Ex-res}} \\
\hline
\hline
     Alice &         51 &     F &      12030 & MAM-pos \\
\hline
       Betty &         52 &       F &      12030 &        CX-neg \\
\hline
      Carol &         51 &       F &      12031 &       CX-pos \\
\hline
     Doris &         52 &     F &      12031 & BS-neg \\
\hline
\end{tabular}}
\hspace{1cm}
\subtable[Generalized transaction data: 1st release]{\label{tbl:gen1}
\begin{tabular}{|r|r|r|r|r|r|}
\hline
 {\bf QI-group} &  {\bf Age} & {\bf Gender} &  {\bf Zip} & {\it {\bf Ex-res}} \\
\hline
\hline
 1 & [51,52] & F & 12030 & MAM-pos \\
\hline
 1 & [51,52] & F & 12030 & CX-neg \\
\hline
 2 & [51,52] & F & 12031 & CX-pos \\
\hline
 2 & [51,52] & F & 12031 & BS-neg \\
\hline
\end{tabular}}\end{center}
\begin{center}
\subtable[Original transaction data at time $\tau_2$]{\label{tbl:or2}
\begin{tabular}{|r|r|r|r|r|}
\hline
{\bf Name} &  {\bf Age} & {\bf Gender} &  {\bf Zip} & {\it {\bf Ex-res}} \\
\hline
\hline
      Alice &         51 &       F &      12030 &       BCM-pos \\
\hline
     Carol &         51 &     F &      12031 & PNE-pos \\
\hline
     Elisa &         51 &       F &      12044 &        MAM-neg \\
\hline
     Fran &         51 &     F &      12045 &       CX-neg \\
\hline
     Grace &         51 &     F &      12040 &       CX-pos \\
\hline
\end{tabular}}
\hspace{1.4cm}
\subtable[Generalized transaction data: 2nd release]{\label{tbl:gen2}
\begin{tabular}{|r|r|r|r|r|r|}
\hline
{\bf QI-group} &  {\bf Age} & {\bf Gender} &  {\bf Zip} & {\it {\bf Ex-res}} \\
\hline
\hline
          3 &    51 &  F &  1203* &       BCM-pos \\
\hline
          3 &    51 &  F &  1203* & PNE-pos \\
\hline
          4 &   51 &  F &  1204* & MAM-neg \\
\hline
          4 &   51 &  F &  1204* & CX-neg \\
\hline
          4 &   51 &  F &  1204* & CX-pos \\
\hline
\end{tabular}}
\end{center}
\end{footnotesize}
\end{table*}

Related work has either focused on anonymization techniques 
dealing with multiple data releases, or on privacy protection techniques
taking into account background knowledge, but limited to a single data
release. We are not aware of any work
taking into account the combination of these conditions. This case
cannot be addressed by simply combining the two types of techniques
mentioned above, since background knowledge can enable new kinds of
privacy threats on sequential data releases.
Extensions of data anonymization techniques to deal with multiple data
releases have been proposed under different 
assumptions~\cite{XiaoSIGMOD07,PeiEDBT08,LiOW08,ZHPJ-EDBT09,BuVLDB08,WongICDE10}.
The work that is closest to ours is probably the one presented
in~\cite{BuVLDB08}, in which sensitive values are divided in
\emph{transient} values that may freely change with time, 
and \emph{persistent} values that never change. 
However, the proposed technique is effective only when
the transition probability among transient values is uniform, and this 
is often not the case, with the medical domain being a clear counterexample. 
In~\cite{WongICDE10} a technique is proposed to defend against
attacks based on the observation of serial data having transient sensitive values; 
however, background knowledge on transition probabilities 
is not considered in that work.
On the contrary, our privacy preserving technique captures non-uniform %
transition probabilities. Our running example in Section~\ref{sec:motivating}
shows that the anonymizations proposed in  
related works are not effective 
when an adversary can obtain background knowledge on the transition probabilities. 
Techniques considering background knowledge have also been
proposed, and they can
be classified according to two main categories: \emph{a)} models based
on logic assertions and rules \cite{Li08}; and \emph{b)} models based on 
probabilistic tools \cite{DuTZ08,LiICDE09}. However, these techniques are
devised for a single release of the data, and, as it is shown in
Section~\ref{sec:exp}, they are ineffective when an adversary having
background knowledge on sequences of sensitive values may observe multiple releases.\\

In this paper we formally model privacy attacks based on background knowledge
extended to serial data releases. 
We present a new probabilistic defense technique taking into account possible 
adversary's background knowledge and how he can revise it each time new data are released.
Similarly to other anonymization techniques, our method is based on the generalization of quasi-identifier (QI) attributes, but generalization is performed  with a new goal: minimizing the difference among sensitive values probability distributions within each QI-group, while considering the knowledge revision process. 
Jensen-Shannon divergence is used as a measure of similarity.
We consider different methods and accuracy levels for the extraction of background knowledge,
and we show that this defense is effective under different combinations of the knowledge of 
the adversary and the defender. 

\textbf{Contributions and paper outline.}
The contributions of this paper can be summarized as follows:
\\
(i) We model privacy attacks on sequential data release based on 
background knowledge about 
the probability distributions of sensitive 
values and sequences of sensitive values. 
We show that  current anonymization techniques are not resistant to these privacy attacks. \\
(ii) We propose \emph{\defense}\ as a new probabilistic defense technique based on Jensen-Shannon divergence.
\\
(iii) Through an experimental evaluation on a large dataset, we show
the effectiveness of our defense under different methods used to extract background knowledge; 
Our results also show that \defense\ provides a very good trade-off between achieved privacy and data utility.

The paper is structured as follows. 
In Section~\ref{sec:motivating}, the privacy problem is presented through an example in the medical domain that illustrates the privacy attacks enabled by  background knowledge, and the inadequacy of state of the art techniques.   
In Section~\ref{sec:bk} we formally model the privacy attack, as well as the considered forms of background knowledge.
In Section~\ref{sec:attack} we show how an adversary can actually extract background knowledge, and revise 
his knowledge in order to perform the attack.
In Section~\ref{sec:defense} we propose our JS-reduce defense algorithm 
that is experimentally evaluated in Section~\ref{sec:exp}. 
Section~\ref{sec:conclusion} concludes the paper.

\section{Motivating scenario}
\label{sec:motivating}
In this section we focus on a specific scenario in the medical domain
to illustrate the privacy attacks enabled by background
knowledge on sequences of sensitive values. The example also shows the inadequacy
of state of the art techniques, and serves as a running example for the
rest of the paper.

\begin{table*}[t!]
\label{tbl:bk}
\caption{Adversary's background knowledge}
\begin{footnotesize}
\begin{center}
\subtable[Sensitive values background knowledge at $\tau_1$]{\label{tbl:bksv}
\begin{tabular}{|r||r|r|r|r|r|}
\hline
\textbf{\emph{Name}} & {\bf Age} &  {\bf Gender} & {\bf Zip} &  {\bf Ex-res} & {\it {\bf $\bksv$}} \\
\hline
\hline
  \emph{Alice} & 51 & F & 12030 & MAM-pos & 0.002\\
\hline
  \emph{Betty} & 52 & F & 12030 & MAM-pos & 0.002\\
\hline\hline
  \emph{Alice} & 51 & F & 12030 & CX-neg & 0.05\\
\hline
  \emph{Betty} & 52 & F & 12030 & CX-neg & 0.05\\
\hline\hline
  \emph{Carol} & 51 & F & 12031 & CX-pos & 0.0003\\
\hline
  \emph{Doris} & 52 & F & 12031 & CX-pos & 0.0003\\
\hline\hline
  \emph{Carol} & 51 & F & 12031 & BS-neg & 0.2\\
\hline
  \emph{Doris} & 52 & F & 12031 & BS-neg & 0.2\\
\hline\hline
  \emph{Alice} & 51 & F & 12030 & BCM-pos & 0.001\\
\hline
\end{tabular}}
\hspace{1.25cm}
\subtable[Sequential background knowledge]{\label{tbl:bkseq}
\begin{tabular}{|r|r|r|r|}
\hline
{\bf Ex-res at $\tau_1$} & {\bf Ex-res at $\tau_2$} & {\bf $\widetilde{p}(s_{\tau_{2}}|s_{\tau_{1}})$} \\
\hline\hline
MAM-pos & BCM-pos & 0.6 \\
\hline
CX-neg & BCM-pos & 0.02 \\
\hline
CX-pos & BCM-pos & 0.02 \\
\hline
BS-neg & BCM-pos & 0.02 \\
\hline
MAM-pos & PNE-pos & 0.02 \\
\hline
CX-neg & PNE-pos & 0.08 \\
\hline
CX-pos & PNE-pos & 0.6 \\
\hline
BS-neg & PNE-pos & 0.02 \\
\hline
\end{tabular}}
\end{center}
\end{footnotesize}
\end{table*}

We consider the case of transaction data representing the results of
medical exams taken by patients, and the need to periodically release
these transactions for data analysis\footnote{We consider analysis
  that require individual transactions; i.e., no aggregation is
  allowed.}.
Each released view contains one tuple for each patient who performed an
exam during the week preceding the publication. We assume that data are
published weekly. For the sake of simplicity, we also assume that
each user cannot perform more than one exam per week; hence, 
no more than one tuple per user can appear in the same view. 
Each generalized tuple includes the age, gender and zip code of the patient, 
as well as the performed exam together with its result. We refer to this latter data, represented
by the multivalue attribute \emph{Ex-res}, as \emph{exam result}\footnote{MAM = mammography, CX = chest 
X\mbox{-}ray, BCM = breast cancer marker, PNE = pneumonia}. 
We denote as positive (\emph{pos}) a result that reveals something anomalous; 
negative (\emph{neg}) otherwise.
The attribute Ex-res is considered the \emph{sensitive attribute}, while
the other attributes play the role of \emph{quasi-identifiers}
(\emph{\qi}), since 
they may be used, joined with external information, to restrict the set
of candidate respondents.
We consider the 
case in which the adversary's background 
knowledge includes both \emph{sensitive values background knowledge} ($\bksv$)
and \emph{sequential background knowledge} ($\bkseq$).
Intuitively, $\bksv$ regards the probability of performing an exam
with a given result based on data such as patient's gender, age, and
ZIP code; e.g., ``middle-aged females have a sensible probability to
undergo a mammography with a positive result (MAM-pos), while
teenagers do not''.  $\bkseq$ regards the probability of a patient's
exam result given the previous exam results.  For instance, ``when the
mammography signals a possible malignancy (MAM-pos) for patient $r$,
there is high probability that a blood sample of $r$ examined within a
month would detect a breast cancer marker (BCM-pos)''.
A simple form of $\bkseq$ is reported in 
Table~\ref{tbl:bkseq}; in
particular, the first row in the table represents the above statement,
where the probability of the event is set to $0.6$. 
As we show in Section~\ref{subsec:extr-bk}, both sequential and 
sensitive values background knowledge can be easily
acquired, either through the scientific literature or from the data. 
We name \emph{posterior knowledge} ($\pksv$) \emph{at $\tau_i$} the adversary's 
confidence about the exam results of tuples respondents after observing the data
released at time $\tau_i$ (e.g., ``The probability that Alice is the respondent of a tuple with
Ex-res = MAM-pos released at $\tau_1$ is $0.5$'').

Consider the original transaction data at time $\tau_1$ (first week)
and $\tau_2$ (second week) shown in Tables~\ref{tbl:or1} and
\ref{tbl:or2}, respectively, and the corresponding generalized
transaction data in Tables~\ref{tbl:gen1} and \ref{tbl:gen2}.
Note that these generalized views satisfy
state of the art techniques for privacy preservation. In particular,
they satisfy $l$-diversity \cite{MachanavajjhalaGKV06} with $l=2$,  $m$-invariance \cite{XiaoSIGMOD07} with $m=2$, as well as the privacy properties proposed in~\cite{ZHPJ-EDBT09,BuVLDB08,RiboniSSDBM09}.
However, we show that the release of these views can lead to a serious privacy threat.
Consider tuples released at $\tau_1$ belonging to QI-group $1$, having
private values MAM-pos and CX-neg, whose possible respondents are 
Alice and Betty. Since Alice and Betty are almost the same age, and
live in the same area, the adversary cannot exploit $\bksv$ 
(reported in Table~\ref{tbl:bksv}) to infer whether Alice or Betty 
is the respondent of the
tuple with private value MAM-pos. 
Hence, his posterior knowledge 
after having observed tuples released at $\tau_1$  states that, both for
Alice and Betty, the probability of being the respondent of one tuple with 
private value MAM-pos is the same of being the respondent of one tuple with 
private value CX-neg, i.e., $0.5$.
Analogously, Carol and Doris have equal probability of being the 
respondent of one tuple with private value CX-pos and 
of one 
with private value BS-neg.

Now, consider tuples released at $\tau_2$ (in Table~\ref{tbl:gen2})
belonging to QI-group $3$, having
private values BCM-pos and PNE-pos, whose possible respondents are 
Alice and Carol. Since Alice and Carol are the same age, and live in very close 
areas, once again the adversary cannot exploit $\bksv$
to infer whether Alice's private value is BCM-pos and Carol's one is PNE-pos, or
vice-versa.
However, the adversary may exploit $\pksv$ at $\tau_1$ and $\bkseq$
to derive a new kind of knowledge, which we name
\emph{revised sensitive values background knowledge} ($\rbksv$) at $\tau_2$.
This knowledge represents the revision of sensitive values background knowledge
computed based on the history of released views, and on sequential background 
knowledge.
The actual method for computing $\rbksv$ is shown in Section~\ref{sec:attack};
here we give an intuition of the adversary reasoning. 
Since the exam result of Alice at $\tau_1$ is either MAM-pos or CX-neg, 
and the one at $\tau_2$ is either BCM-pos or PNE-pos, $4$ possible
sequences of sensitive values about Alice exist. Among these sequences, according to
$\bkseq$, the one having MAM-pos at $\tau_1$ and BCM-pos at $\tau_2$ is more 
probable than the others, since a positive mammography result is frequently
followed by a positive breast cancer marker test.
Analogously, among the possible sequences regarding Carol, the most
probable is the one having CX-pos at $\tau_1$ and PNE-pos at $\tau_2$. 
Through this kind of reasoning the adversary revises his sensitive
values background knowledge, associating high confidence to the fact that at 
$\tau_2$ Alice is positive to breast cancer markers, while Carol has pneumonia. 
Hence, based on $\rbksv$, 
the adversary can assign with high confidence the correct sensitive values to Alice and Carol.

%
%
%
%
%
%
%

%

 
\section{Modelling attacks based on background and revised knowledge}
\label{sec:bk}
In this section we formally model privacy attacks based on background and revised knowledge available to an adversary. 
\subsection{Problem definition}
\label{sec:preliminaries}
We denote by $V_i$ a view on the original transaction data at time $\tau_i$, and
by $V^*_i$ the generalization of $V_i$ released by the data publisher. 
We denote by 
$\displaystyle{\mathcal{H}^{*}_{j} = \langle V_1^*, V_{2}^*, \ldots, V_j^* \rangle}$ 
a \emph{history} of released generalized views.
We assume that the schema remains
unchanged throughout the release history, and we partition the view columns into
a set $A^{qi} = \{A_1, A_2, \ldots, A_m\}$ of quasi-identifier attributes, 
and into a single private attribute $S$. For the sake of simplicity, we 
assume that the domain of each quasi-identifier attribute is numeric, but
our notions and techniques can be easily extended to categorical attributes. 
Given a tuple $t$ in a view and an attribute $A$ in its schema, $t[A]$ is the projection of tuple $t$ onto $A$.

Views are generalized by a \emph{generalization function} $G()$ that removes possible 
explicit identifiers from the original tuples, and generalizes the quasi-identifiers.
Tuples in $V^*_j$ are partitioned into \emph{QI-groups}; i.e., sets of tuples 
having the same values for their quasi-identifier attributes. 
Even if we consider generalization-based anonymity, both our attack model and defense method
can be seamlessly applied to bucketization-based techniques.

At each release of a view $V^*_j$, the goal of an adversary is to reconstruct, 
with a certain degree of confidence, the 
\emph{sensitive association} between the identity of a respondent of a tuple $t$ in
$V^*_j$ and her sensitive value $t[S]$.
The adversary model considered in this paper is based on the following
assumptions:
\squishlist
\item The generalization function $G()$ is publicly known.
\item The adversary may have external information about respondents' 
personal data. For example, for each \QIG\ $Q$, the adversary may know its
set of respondents.
\item The adversary may observe a history $\Hi^*_j$ of anonymized views.
\item The adversary may have background knowledge on sensitive values $\bksv$ and $\bkseq$ as formally defined in 
Sections~\ref{subsec:bksv} and~\ref{subsec:bkseq}, respectively.
\squishend
Note that the first two assumptions are shared by most work on anonymity. 
As illustrated in Section~\ref{sec:introduction}, the third and the fourth 
(limited to $\bksv$) have also been considered by related work  but not in 
combination. Finally,  $\bkseq$ is original to this work. 

\subsection{Sensitive values background knowledge ($\bksv$)}
\label{subsec:bksv}
Sensitive values background knowledge %
represents the a-priori probability 
of associating an individual to a sensitive value. 
$\bksv$ is modeled according to the following definition.

\begin{definition}\label{def:bksv}
The \emph{sensitive values background knowledge} is a function 
$\bksv : R \rightarrow \Upsilon$, where 
$R$ is the set of possible respondents' identities, and 
\[
   \Upsilon = \{ (p_1, \ldots, p_n) \;
   | \sum_{1 \leq i \leq n}p_i = 1 \; (0 \leq p_i \leq 1)\}
\]
is the set of
possible probability distributions of $S$, where 
$D[S] = \{s_1, s_2, \ldots, s_n\}$.
\end{definition}

For example, if $r\in R$ is a possible respondent of a tuple in a released view, $\bksv(r)$ returns, for each sensitive value $s_j \in D[S]$, the probability $p_j$ of $r$ being actually associated with $s_j$.

\subsection{Sequential background knowledge ($\bkseq$)}
\label{subsec:bkseq}
We model the sensitive value referring to a respondent $r$ %
by means of the discrete random variable $\mathcal{S}$ having values in $D[S]$. 
Hence, sequential background knowledge is a function 
that returns the probability distribution of $\mathcal{S}$ at $\tau_j$ 
given a sequence 
$\Lambda = \langle s_1, s_2, \ldots, s_{j-1} \rangle$
of past observations at 
$T = \langle \tau_1, \tau_2, \ldots, \tau_{j-1} \rangle$.

\begin{definition}\label{def:bkseq}
\emph{The sequential background knowledge} is a function 
$\displaystyle{\bkseq: \overline{\Lambda} \times \overline{T} \times R \times \mathcal{T} \rightarrow \Upsilon}$,
where 
$\overline{\Lambda}$ is the set of possible sequences 
of past observations of a respondent's sensitive values, $\overline{T}$ is the 
set of possible sequences of time instants at which the observations were taken, 
$R$ is the set of respondents' identities,
$\mathcal{T}$ is the set of possible time instants, and 
$\Upsilon$ 
is the set of
possible probability distributions of $\mathcal{S}$.
\end{definition}

For example, if $r\in R$ is a possible respondent of a tuple in a released view, and the adversary knows that $r$ has been associated with values $s_1$, and $s_2$ at past instants $\tau_1$, $\tau_2$, respectively, then $\bkseq$ returns the probability $p_j$ of $r$ being associated with $s_j$ at $\tau_3$, for each possible sensitive value $s_j$.  

\begin{figure}[t!]
  \centering
\begin{minipage}[t]{\columnwidth}
\centering 
 \includegraphics[width=\columnwidth]{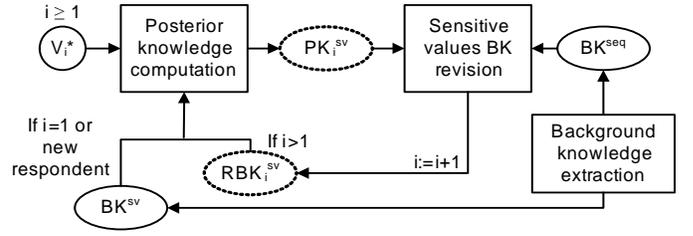}
\end{minipage}
  \caption{Adversary's inference mechanisms}
  \label{fig:inference-attack}
\end{figure}

\subsection{Posterior ($\pksv$) and revised sensitive values background knowledge ($\rbksv$)}
\label{subsec:pksv}
As intuitively described in the running example of Section~\ref{sec:motivating}, posterior knowledge at $\tau_i$ represents the adversary's confidence 
about the association between a respondent and sensitive values \emph{after} the observation
of view $V^*_{i}$.
For the sake of readability, we denote $\pksv$ \emph{at $\tau_i$} by $\pksv_i$.
\begin{definition}\label{def:pksv}
The \emph{posterior knowledge} is a function 
$\pksv : R \times \mathcal{T} \rightarrow \Upsilon$, where 
$R$ is the set of respondents' identities, $\mathcal{T}$ is the set of possible
time instants, and 
$\displaystyle{\Upsilon}$ %
is the set of possible probability distributions of $\mathcal{S}$. 
\end{definition}

A method to compute $\pksv$ is described in Section~\ref{subsec:comp-pksv}.\\

After observing view $V^*_{j-1}$, an adversary may exploit 
posterior knowledge at $\tau_1$, $\tau_2$, $\ldots$, $\tau_{j-1}$, 
together with sequential background knowledge $\bkseq$, to derive 
new information about the probability distribution of $\mathcal{S}$
at $\tau_j$. %
We call this information \emph{revised sensitive values background knowledge}
at $\tau_j$ (denoted as $\rbksv_j$); it is essentially the revision
of sensitive values background knowledge due to the observation of a history
of released tuples. 
$\rbksv_j$ can be used by an adversary to calculate posterior knowledge
after the observation of $V^*_j$.

The \emph{revised sensitive values background knowledge} is a function 
$\displaystyle{\rbksv}$ 
having the same domain and co-domain as function $\displaystyle{\pksv}$ defined in
Definition~\ref{def:pksv}. 
The method to compute $\rbksv$ is described in Section~\ref{subsec:comp-rbksv}.

\subsection{The privacy attack}
\label{subsec:privacy-attack}
The inference method adopted by an adversary to reconstruct the
sensitive association is depicted in
Figure~\ref{fig:inference-attack}. 
The adversary obtains sensitive values 
background knowledge $\bksv$, as well as sequential background knowledge $\bkseq$,
using one of the techniques explained in Section~\ref{subsec:extr-bk}. 
When the first view $V^*_1$ is released at time $\tau_1$, the adversary 
computes posterior knowledge $\pksv_1$ based on
$V^*_1$ and on $\bksv$; a method for posterior knowledge computation is
presented in Section~\ref{subsec:comp-pksv}.
Then, the adversary computes revised sensitive values background 
knowledge $\rbksv_2$, based on $\pksv_1$ and on sequential background 
knowledge $\bkseq$. A technique for knowledge revision is illustrated in 
Section~\ref{subsec:comp-rbksv}.
Hence, when view $V^*_2$ is released, the adversary 
computes $\pksv_2$ based on $V^*_2$ and on
$\rbksv_2$.
Then, the knowledge revision cycle
continues with the computation of $\rbksv_3$ based on
$\pksv_2$ and $\bkseq$, and so on.  
When $V^*_i$ includes a tuple of respondent $r$, and no
tuples of $r$ appeared in $\mathcal{H}^*_{i-1}$, 
$\rbksv(r,\tau_i)$ cannot be computed, since no historical information 
about $r$'s tuples is available; in this case $\bksv$ is used instead of $\rbksv(r,\tau_i)$.


\section{Knowledge extraction and revision}
\label{sec:attack}
In this section we illustrate how an adversary may obtain background
knowledge, and use it to reconstruct the association between
respondents of released tuples and their sensitive values.

\subsection{Extracting background knowledge}
\label{subsec:extr-bk}
Intuitively, the more accurate is the adversary's background knowledge
(i.e., close to the underlying process that generated the data), the more effective will be his 
attack. Background knowledge can be obtained using different 
methods, depending on the available data, and on the data domain. 

The problem of extracting sensitive values background knowledge based on a corpus
of available data has been thoroughly studied, 
and effective techniques are 
available (e.g., the ones proposed in~\cite{Li08,DuTZ08,LiICDE09}). 
Hence, in the rest of this paper we assume that the adversary extracts $\bksv$
using one of the existing methods. 
However, existing privacy-preserving techniques do not consider
the extraction of $\bkseq$. For this reason, we illustrate how this
knowledge can actually be obtained.

\squishlist
\item \emph{Incrementally extracting $\bkseq$ from the data to be released}. 
One of the methods proposed to compute the background knowledge that an adversary 
may obtain is to extract it from the same data that are going to be generalized 
and released~\cite{Li08,LiICDE09}. 
At the time of writing, these techniques are limited to the calculation of 
$\bksv$. However, based on a sequence $\mathcal{H}_i$ of 
original views, sequential pattern mining (SPM) methods~\cite{AgrawalICDE95}
can be used to calculate a function $\iebkseq$ 
that approximates the exact $\bkseq$. That function is incrementally refined as
long as new original views are available. 
A number of different SPM techniques have been proposed in the last years for 
different application domains (e.g.,~\cite{SpmKDD02,SpmCIKM04,SpmTKDE04},
among many others). 
Hence, the choice of the most appropriate SPM algorithm strongly depends on 
the domain of the data.  
In Section~\ref{subsec:exp-bkext} we illustrate the algorithm we adopt to calculate 
$\iebkseq$ for the sake of our experiments.
Of course, this technique can be used by the defender only, since we assume that 
the adversary cannot observe original views. 
\item \emph{Mining $\bkseq$ from an available corpus of data}. 
Even if an adversary cannot observe the original data, he may apply
SPM methods to a corpus of external data from the same domain to calculate a function $\spmbkseq$ that
approximates the exact $\bkseq$.
\item \emph{Exploiting domain knowledge}. In many cases it is possible to exploit domain knowledge 
extracted from the scientific literature. For instance, in the medical domain, a number of
surveys have been published, which report accurate statistics about the probability
of disease evolution with time (e.g.,~\cite{Sepsi,hiv,Hepatitis,Alzhaimer}, just 
to name a few). Given this knowledge, it is easy to 
design a function
$\dkbkseq$, which approximates the exact $\bkseq$.
\squishend

\subsection{Computing posterior knowledge}
\label{subsec:comp-pksv}
In order to compute $\pksv_i$, 
it is possible to reason considering a QI-group at a time.
In particular, in our case, given a QI-group $Q$ having $R$ as the set of respondents, a 
\emph{possible configuration} is a function $c : \mathcal{Q} \rightarrow \mathcal{R}$,
i.e., a one-to-one correspondence between elements in $Q \in \mathcal{Q}$ and elements in $R \in \mathcal{R}$. 
Given a possible configuration $c$, for each tuple $t \in Q$ we say that ``$r$ is the respondent
of $t$ in the possible configuration $c$'' if $c(t)=r$.

 \begin{example}\label{ex:possible-configs}
 Consider Table~\ref{tbl:gen2} released at $\tau_2$ in our running example, and 
 QI-group $3$ composed of Alice's and Carol's tuples. In this case, two possible
 configurations $c_1$ and $c_2$ exist. According to $c_1$, Alice is the respondent of the 
 tuple with sensitive value BCM-pos, and Carol is the respondent of the one with
 PNE-pos. According to $c_2$, Alice is the respondent of the tuple with PNE-pos,
 and Carol is the respondent of the one with BCM-pos.
 \end{example}

 Each possible configuration $c_j$ is associated to a
 confidence degree $d_j$, that depends on the background knowledge of
 the adversary. $d_j$ is computed as the sum of the
 probabilities, given by $\rbksv$ (or $\bksv$), of the single
 associations between respondents and sensitive values in $c_j$.

Given $r \in R$, and the set $C$ of possible configurations, in order to
calculate $\pksv(r,\tau_i) = (p_1, p_2, \ldots, p_n)$ we need to compute, for each
$p_m \in \{ p_1, p_2, \ldots, p_n \}$, the sum of the degree of confidence of every
possible configuration in which $r$ is the respondent of a tuple having sensitive value
$s_m$, divided by the sum of the degree of confidence of every possible configuration: 
\[
   p_m = \frac{{\sum_{\forall c_j \in C : \, c_j(t)=r \wedge t[S]=s_m} d_j}}
              {{\sum_{\forall c_j \in C} d_j}}.
\]

 \begin{example}
   Continuing Example~\ref{ex:possible-configs}, according to
   $\rbksv_2$ (Table~\ref{tbl:kpos1b}), the degree of confidence for
   $c_1$ is much higher than the one for $c_2$.  Indeed, the
   probability of Alice being the respondent of a tuple with sensitive
   value BCM-pos is $0.31$, which is also the probability of Carol
   being the respondent of the other tuple; hence, $d_1 = 0.31 + 0.31
   = 0.62$ .  The probabilities regarding configuration $c_2$ are much
   lower; i.e., $0.05$ and $0.02$, respectively; i.e., $d_2 = 0.07$.
   Hence, if $p_m$ is the probability of Alice being the respondent of
   a tuple with sensitive value BCM-pos, by applying the above formula
   we obtain $p_m = \frac{0.62}{0.62 + 0.07} \simeq 0.9$.  The values
   of $\pksv$ at $\tau_2$ are shown in Table~\ref{tbl:kpos2}.
 \end{example}

However, in general the exact computation of $\pksv$ is intractable;
indeed, if the cardinality of the QI-group is
$k$, the number of possible configurations is $k!$.
For this reason, an approximate algorithm
is the natural candidate for the computation of posterior knowledge. 
In our experimental evaluation, we calculate posterior knowledge by the 
$\Omega$-estimate method proposed by Li et al. \cite{LiICDE09}.
\begin{table}[t!]
\label{tbl:ik}
\caption{Adversary's posterior and revised knowledge}
\begin{footnotesize}
\begin{center}
\begin{tabular}{l p{0.6\columnwidth}}
\subtable[$\pksv$ at $\tau_1$]{\label{tbl:kpos1}
\begin{tabular}{|r|r|r|r|}
\hline
{\bf Name} & {\bf Ex-res} & {\it {\bf $p$}} \\
\hline
\hline
  Alice & MAM-pos & 0.5\\
\hline
  Alice & CX-neg & 0.5\\
\hline\hline
  Betty & MAM-pos & 0.5\\
\hline
  Betty & CX-neg & 0.5\\
\hline
\hline
  Carol & CX-pos & 0.5\\
\hline
  Carol & BS-neg & 0.5\\
\hline\hline
  Doris & CX-pos & 0.5\\
\hline
  Doris & BS-neg & 0.5\\
\hline
\end{tabular}} &
\subtable[$\rbksv$ at $\tau_2$]{\label{tbl:kpos1b}
\begin{tabular}{|r|r|r|} %
\hline
{\bf Name} & {\bf BCM-pos} & {\bf PNE-pos} \\ %
\hline
Alice & 0.31 & 0.05 \\ %
\hline
Carol & 0.02 & 0.31 \\ %
\hline
\end{tabular}} %
\subtable[$\pksv$ at $\tau_2$]{\label{tbl:kpos2}
\begin{tabular}{|r|r|r|r|r|}
\hline
{\bf Name} & {\bf Ex-res} & {\it {\bf $p$}} \\
\hline
\hline
  Alice & BCM-pos & 0.9\\
\hline
  Alice & PNE-pos & 0.1\\
\hline\hline
  Carol & BCM-pos & 0.1\\
\hline
  Carol & PNE-pos & 0.9\\
\hline
\end{tabular}} 
\end{tabular}
\end{center}
\end{footnotesize}
\end{table}

\subsection{Computing revised knowledge}
\label{subsec:comp-rbksv}
In order to compute revised sensitive values background knowledge at $\tau_i$ 
($i > 1$) 
the adversary needs to
calculate, for each respondent $r$ of a tuple in $V^*_i$, and for each
sensitive value $s \in D[S]$, the marginal probability of $r$ to be
the respondent of a tuple with private value $s$ in $V^*_i$, given
$\pksv$ and $\bkseq$.  Let $\mathcal{V}^* = \langle V^*_1, V^*_2,
\ldots, V^*_{i-1} \rangle$ be the history of released views containing
a tuple of $r$, and $\mathcal{S}_i$ the random variable representing
the sensitive value of $r$'s tuple released at $\tau_i$.  Then, by
applying the conditioning rule, we have:
\[
   P(\mathcal{S}_i) = 
   \sum_{\lambda \in \Lambda}\Big(\bkseq(\lambda,T,r,\tau_i) \cdot P(\lambda)\Big),
\]
where $T = \langle \tau_1, \tau_2, \ldots, \tau_{i-1} \rangle$,
$\Lambda$ is the set of possible sequences of sensitive values of $r$'s tuples
released at $T$, and $P(\lambda)$ is the probability of sequence $\lambda \in \Lambda$.
In particular, given the sequence $\lambda = \langle s_1, s_2, \ldots, s_{i-1} \rangle$,
$P(\lambda)$ is the joint probability of the occurrence
of each $s_j \in \lambda$ at $\tau_j$ %
based on $\pksv$. If we denote as $p(r, s_j, \tau_j)$ that probability according
to $\pksv(r,\tau_j)$, we have:
\[
   P(\lambda) = 
   \prod_{s_j \in \lambda}\Big( p(r, s_j, \tau_j) \Big).
\]

\begin{example}
 Considering  our running example, the adversary
 revises his sensitive values background knowledge after observing
 view $V^*_1$ to obtain $\rbksv_2$ as follows.
The probability $p(\textrm{\it Alice}, s, \tau_1)$ 
 that Alice is the respondent of a tuple released at $\tau_1$ having sensitive 
 value $s$ is given by $\pksv_1$ (Table~\ref{tbl:kpos1}). 
 Moreover, we represent by $\widetilde{p}(\textrm{\it BCM-pos} \, | \, s)$ the probability 
 that an individual is the respondent of a tuple released at $\tau_2$ with sensitive 
 value BCM-pos provided that the same individual was the respondent of a tuple
 released at $\tau_1$ with sensitive value $s$; this conditional 
 probability is given by $\bkseq$ (Table~\ref{tbl:bkseq}). 
 Then, the marginal 
 probability of Alice to be the respondent of one tuple with %
 BCM-pos at $\tau_2$ can be calculated as: 
 \begin{align*}
 	p(\textrm{\it Alice},& \textrm{\it BCM-pos}, \tau_2) = \\
 	&= \sum_{\forall s \in D[S]}\Big(p(\textrm{\it Alice}, s, \tau_1) \cdot 
 	\widetilde{p}(\textrm{\it BCM-pos} \, | \, s) \Big) = \\
 	&= p(\textrm{\it Alice}, \textrm{\it MAM-pos}, \tau_1) \cdot 
 	\widetilde{p}(\textrm{\it BCM-pos} \, | \textrm{\it MAM-pos}) + \\
 	&+ p(\textrm{\it Alice}, \textrm{\it CX-neg}, \tau_1) \cdot 
 	\widetilde{p}(\textrm{\it BCM-pos} | \textrm{\it CX-neg})= \\
 	&= 0.5 \cdot 0.6 + 0.5 \cdot 0.02 = 0.31.
 \end{align*} 

 Conditioning over any possible private value $s'$ other than MAM-pos
 and CX-neg is omitted from the above formula, since the probability
 $p(\textrm{\it Alice}, s', \tau_1)$ according to $\pksv_1$ is $0$.
 Analogously, the adversary calculates that, according to $\rbksv_2$,
 Alice has $0.05$ probability to be the respondent of a tuple with
 private value PNE-pos, while the probability of Carol is $0.31$ for
 PNE-pos, and $0.02$ for BCM-pos (Table~\ref{tbl:kpos1b}).
 \end{example}


\section{JS-reduce defense}
\label{sec:defense}
In this section we illustrate the \emph{JS-reduce} defense
against the identified background knowledge attacks.\\

\subsection{Defense strategy} 
In order to enforce anonymity, it is
necessary to limit the adversary's capability of identifying the actual
respondent of a tuple in a given \QIG. Referring to the terminology introduced in 
Section~\ref{subsec:comp-pksv} and to the attack we are considering, this means 
reducing the confidence of the adversary 
in discriminating a configuration $\widetilde{c}$ among the possible ones, based on his 
knowledge $\rbksv$.

The goal of \defense\ is to create \QIG s whose tuple respondents have
similar $\rbksv$ ($\bksv$) distributions.  Indeed, if the respondents
of tuples in a \QIG\ are indistinguishable with respect to $\rbksv$
($\bksv$), the adversary cannot exploit background knowledge to
perform the attack.  Of course, defending against background knowledge
attacks is not sufficient to guarantee privacy protection against
other kinds of attacks.  For this reason, \defense\ also enforces
$k$-anonymity and $t$-closeness, in order to protect against
well-known identity- and attribute-disclosure attacks,
respectively. Note that \defense\ can be easily extended to enforce
additional privacy models.

\subsection{Defending against sequential background knowledge attacks} 
\label{subsec:def-bkseq}
In order to measure the similarity of probability distributions
$\rbksv$ ($\bksv$), we adopt \emph{Jensen-Shannon divergence
  (JS)}~\cite{lin91}.  With respect to other distance measures among
probability distributions, this function has three important
properties: \emph{i)} it can be computed on a set of more than two
distributions; \emph{ii)} it is always a definite number; \emph{iii)}
it is symmetric with respect to the order of the arguments. Suppose
that $\mathbf{P} = \{ \overline{p}^1, \ldots, \overline{p}^u\}$ is a
set of probability distributions such that each element has form:
$\overline{p}^i = (p_1^i, \ldots, p_s^i)$. Suppose also that $\pi^1,
\ldots, \pi^u$ denote the \emph{weights} of the probability
distributions, and that $\sum_{i = 1}^u{\pi^i} = 1$.  Then the JS
divergence among distributions in $\mathbf{P}$ is:
\[\textrm{\emph{JS}}(\mathbf{P}) = H(\sum_{i = 1}^u{\pi^i \cdot \overline{p}^i}) -
\sum_{i = 1}^u{\pi^i \cdot H(\overline{p}^i)},\]
where $H(\overline{p})$ is the Shannon entropy of $\overline{p} =
(p_1, \ldots, p_s)$.  In our case, each $\overline{p}^i$ corresponds
to the background knowledge about a tuple respondent; since this
probability $\overline{p}^i$ already includes the adversary's
confidence, when we compute the above formula we assign the same
weight to each probability distribution.

Given a required threshold $j$, the \defense\ defense guarantees that, 
for each \QIG\ $Q$ in an anonymized view, the JS divergence
of the set of probability distributions $\rbksv$ ($\bksv$) of respondents
of tuples in $Q$ is below $j$. 
Note that, given the privacy preferences expressed by the data owner,
the actual value of threshold $j$ must be chosen according to many
domain-specific factors, including the diversity of sensitive values
in released views, and background knowledge.  Similar considerations
apply for the choice of the parameter $k$ of $k$-anonymity and $t$ of
$t$-closeness.

\begin{figure}[t!]
  \centering
\begin{minipage}[t]{\columnwidth}
\centering
  \includegraphics[width=\columnwidth]{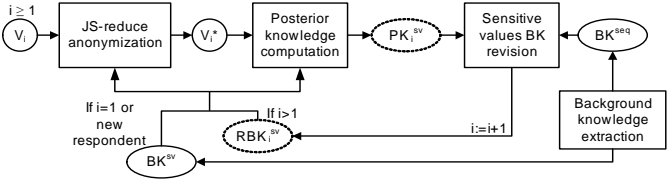}
\end{minipage}
  \caption{Defense mechanisms}
  \label{fig:inference-defense}
\end{figure}

Clearly, in order to be effective against sequential background knowledge 
attacks, \defense\ needs to calculate the
$\rbksv$ distribution of respondents before anonymizing data.
Hence, similarly to the knowledge revision cycle presented in
Section~\ref{sec:attack}, the defense technique 
(graphically illustrated in Figure~\ref{fig:inference-defense}), 
performs posterior 
knowledge computation, and sensitive values background knowledge revision. 
$\bksv$ and $\bkseq$ are obtained using one of the techniques illustrated in
Section~\ref{subsec:extr-bk}.

\SetAlFnt{\footnotesize}
\linesnumbered 
\begin{algorithm}[t!]
\caption{\defense\ algorithm}
\label{alg:general-defense}

 \dontprintsemicolon

 \KwIn{Sequence $\mathcal{H}_n = \langle V_{1}, \ldots, V_{n} \rangle$,
 the set $R$ of possible respondents as well as their \qi\ values,
 $\bksv$, $\bkseq$, 
 the minimum level $k$ of \ka, threshold $t$ of \tc, threshold $j$ of JS divergence.
 }
 \KwOut{$V_n^*$
 }

\BlankLine

 \textbf{JS-reduce$(\mathcal{H}_n, R, \bksv, \bkseq, k, t, j)$}
\\\Begin{
   \ForAll{$r \in R$}{
     $\rbksv_1(r) \leftarrow \bksv(r)$\;
   }
   \For{$h = 1$ to $n$}{
     $V_h^* \leftarrow$ Generalize$(V_h, RBK_{h}^{sv}, t, j, k)$\;
     \ForAll{$r \in R_h$}{
       $\pksv_h(r) \leftarrow$ PKComputation$(V_h^*, \rbksv_h, r)$\;
       $\rbksv_{h+1}(r) \leftarrow$ BKRevision$(\pksv(r), \bkseq, r)$ \;%
     }
   }
   \Return{$V^*_n$} 
 }

\BlankLine
\BlankLine
\BlankLine
\BlankLine

 \setcounter{AlgoLine}{0}
 \KwIn{The anonymized release $V_h^*$, the set $\rbksv_h$ of revised
   background knowledge for each respondent of a tuple in $V_h^*$, respondent $r$} 

\KwOut{$\pksv_h(r)$ }
 \textbf{PKComputation$(V_h^*, \rbksv_h, r)$}
 \\\Begin{
    QI-group $Q \leftarrow Q' \in V^*_h$ s.t. $r$ is the respondent of one tuple in $Q'$ \;
     C $\leftarrow$ $\{c_j \mid c_j $ is a valid configuration for $Q \}$ \;
     \ForAll{$c_j \in C $}{
       confidence degree $d_j \leftarrow 0$ \;
       \ForAll{$r'$ s.t. $\exists t \in Q \mid c_j(t) = r'$}{
	 $t' \leftarrow t \mid c_j(t) = r'$ \;
         $d_j \leftarrow d_j  + RBK_h^{sv}(r', t'[S])$ \;
       }
     }
     \ForAll{$s \in D[S]$}{
       $p(r, s) \leftarrow \frac{\sum_{\forall c_j \in C | c_j(t) = r \wedge t[S] = s}{d_j}}{\sum_{c_j \in C}{ d_j }} $ \;
     }
     $\pksv_h(r) \leftarrow \left\{ p(r,\tilde{s}), \forall \tilde{s} \in D[S] \right\}$ \;
     \Return{$\pksv_h(r)$} 
}

\BlankLine
\BlankLine
\BlankLine
\BlankLine

 \KwIn{The set of posterior knowledge of respondent $r$ $\pksv(r) = \{
   \pksv_1(r), \dots , \pksv_h(r) \}$, the available sequential
   background knowledge $\bkseq$, respondent $r$}

\KwOut{$\rbksv_{h+1}(r)$ }

\setcounter{AlgoLine}{0}
 \textbf{BKRevision$(\pksv(r), \bkseq, r)$}
 \\\Begin{
   $\Lambda \leftarrow \{ \lambda = \langle s_1, \dots , s_i \rangle \mid s_j$ is a possible sensitive value for $r$ released at $\tau_j\}$ \;
   \ForAll{$\lambda \in \Lambda$}{
     $P(\lambda) \leftarrow 1$ \;
     \ForAll{$s_j \in \lambda$}{
       $P(\lambda) \leftarrow P(\lambda) \cdot \pksv_j(r, s_j)$\;
     }
   }
   \ForAll{$s \in D[S]$}{
    $\widetilde{p}(s \mid \lambda)$ is the conditional probability given by $\bkseq$ \;
    $ p(s) \leftarrow {\sum_{\lambda \in \Lambda}{ \widetilde{p}(s \mid \lambda)} \cdot P(\lambda)}$ \;
   }
   $\rbksv_{h+1}(r) \leftarrow \{ p(s), \forall s \in D[S]\}$ \;
   \Return{$\rbksv_{h+1}(r)$}
}

\end{algorithm}

\subsection{The \defense\ algorithm} 
The pseudo-code of the \defense\ algorithm is shown in
Algorithm~\ref{alg:general-defense}.
The algorithm takes as input: 
\emph{i)} a sequence 
$\mathcal{H}_n = \langle V_{1}, \ldots, V_{n} \rangle$ of original views;
\emph{ii)} the set $R$ of respondents of tuples in $\mathcal{H}_n$,
as well as their \qi\ values; 
\emph{iii)} sensitive values background knowledge  $\bksv$ and sequential background knowledge $\bkseq$;
\emph{iv)} the minimum level $k$ of \ka, threshold $t$ of \tc, and 
threshold $j$ of JS divergence.
It returns $V^{*}_{n}$, the generalization of $V_{n}$.

At first (lines $3$ to $5$), for each respondent of tuples in
$\mathcal{H}_n$, $\rbksv$ at $\tau_1$ is initialized according to
$\bksv$.  Then (lines $5$ to $11$), each view $V_i$ in $\mathcal{H}_n$
is processed in turn, from $V_1$ to $V_n$.  In particular, each $V_i$
is generalized by the \emph{Generalize} procedure (line $6$) in order
to enforce thresholds $j$ of JS divergence, $t$ of $t$-closeness, and
minimum cardinality $k$.  The algorithm for generalization,
specifically designed to preserve the data quality, is described in
detail in Section~\ref{subsec:bucket}.  We call $V^*_i$ the
generalization of $V_i$, and $R_i$ the set of respondents of tuples in
$V^*_i$. After the generalization, for each respondent in $R_i$,
\defense\ calculates the posterior knowledge (line $9$) and the
revised sensitive values background knowledge (line $10$) at
$\tau_{i+1}$.
Finally (line $12$), the generalized view $V^*_n$ is returned.
Procedures \emph{PKComputation} and \emph{BKRevision} apply the
adversary inference mechanisms described in
Section~\ref{subsec:comp-pksv} and Section~\ref{subsec:comp-rbksv},
respectively.
As for other privacy-preserving techniques
(e.g.,~\cite{XiaoSIGMOD07,RiboniSSDBM09}), it is possible that some
tuples cannot be arranged in any \QIG\ without violating some of the
privacy requirements. In this case, JS-reduce suppresses those
tuples. Experimental results, reported in Section~\ref{sec:exp}, show
that the percentage of suppressed tuples is negligible. For those
domains in which suppression of tuples is not acceptable, \defense\
can be easily modified to enforce the required thresholds by the
insertion of counterfeit tuples.

\SetAlFnt{\footnotesize}
\begin{algorithm}[t!]
\caption{Generalization procedure}
\label{alg:anonymization}
 \dontprintsemicolon

\textbf{Generalize$(V_h, t, j, k)$}
\\\Begin{
  $V_h^* = \emptyset$\;
  \ForAll{$v \in V_h$}{
   $i_v \ \leftarrow$ ComputeHilbertIndex($v$)\;
  }
  $\widetilde{V}_h \leftarrow$ OrderOnHilbertIndex($V_h$)\;
  $Q \leftarrow \emptyset$ \;
  \For{$\tilde{v} = v_1$ to $v_{|\widetilde{V}_h |}$}{
    $Q \leftarrow Q \cup {\tilde{v}}$ \;
    \If{$|Q| \geq k \wedge t\mbox{-}clos(Q) \leq t \wedge js(Q) \leq j$}{
       CreateQIG($Q$) \;
       $Q \leftarrow \emptyset$ \;
    }
  }
    \If{$Q \neq \emptyset$}{
    Remove tuples $v \in Q$ \;
  }

  \Return{$V_h^*$}
}

 \BlankLine
 \BlankLine

 \setcounter{AlgoLine}{0}
 \textbf{CreateQIG$(Q)$}
 \\\Begin{
    GeneralizeQIvalues($Q$)\;
    $V_h^* \leftarrow V_h^* \cup Q$ \;
}

\end{algorithm}

\subsection{Data quality-oriented generalization}
\label{subsec:bucket}
Any anonymization technique based on \qi\ generalization needs to
carefully consider the resulting data quality: the more the \qi\
values are generalized, the lower is the quality (and utility) of
released data.
Hence, instead of adopting a general-purpose anonymization framework
such as Mondrian~\cite{LeFevreICDE06}, we devised an ad-hoc \qi\
generalization technique for \defense\ to achieve better data quality.
Note that finding the optimal generalization of data that satisfies
the privacy requirements of \defense\ (i.e., the one that minimizes
\qi\ generalization) is an NP-hard problem; indeed, it is well known
that even optimal $k$-anonymous generalization is
NP-hard~\cite{Meyerson04}.  For this reason, we devised an approximate
algorithm, whose pseudo-code is shown in
Algorithm~\ref{alg:anonymization}. The \emph{Generalize} procedure
receives as input: \emph{i)} the original view $V_h$; \emph{ii)}
revised sensitive values background knowledge at $\tau_h$; \emph{iii)}
a minimum level $k$ of \ka, threshold $t$ of \tc\ and threshold $j$ of
\js.  It returns $V_h^*$, the generalization of $V_h$.

As proposed in~\cite{GhinitaVLDB07},
in order to partition tuples in \QIG s, the procedure exploits the
Hilbert space-filling curves.\footnote{A Hilbert space-filling curve
  is a function that maps a point in a multi-dimensional space into an
  integer. With this technique, two points that are close in the
  multi-dimensional space are also close, with high probability, in
  the one-dimensional space obtained by the Hilbert transformation.}
For each tuple in $V_h$, function \emph{ComputeHilbertIndex} (lines
$4$ to $6$) computes its Hilbert index considering the
multi-dimensional space having the \qi\ attributes as dimensions.
Then, tuples in $V_h$ are re-ordered with respect to their Hilbert
index, obtaining an auxiliary list $\widetilde{V}_h$ (line $7$).
The procedure adds to a group $Q$ a tuple from the ordered
list $\widetilde{V}_h$, and
checks if the cardinality of the group is greater than the \ka \
threshold $k$, and if the \tc \ and \js \ values of that group are
below thresholds $t$ and $j$, respectively.
Note that, according to the Hilbert transformation, tuples with
similar \qi\ values are close in the list $\widetilde{V}_h$, and
respondents having similar \qi\ values are also likely to have similar
probability distributions according to $\bksv$.
Hence, we achieve both of our goals: \emph{i)} it is likely to find
groups of tuples satisfying privacy constraints, and \emph{ii)} we
limit the generalization of \qi\ values.  Then, if the required
privacy constraints are satisfied, a new \QIG\ is created (line $12$)
by procedure \emph{CreateQIG}: the \qi \ values are substituted with
intervals including the \qi\ values of each tuple; the same procedure
is repeated with the remaining tuples. Otherwise (if constraints are
violated), the next tuple in $\widetilde{V}_h$ is added to the group
until the constraints are satisfied (line $10$).

As explained in Section~\ref{sec:defense}, it may happen that a few
tuples cannot be grouped into a \QIG \ (line $16$) during the first
phase.  In the current version of the algorithm, those tuples are
suppressed in order to guarantee the privacy constraints in the whole
view.  However, the algorithm can be easily modified to apply other
solutions; e.g., based on the creation of counterfeit tuples.


\section{Experimental evaluation}
\label{sec:exp}
In this section we present an experimental evaluation of the 
privacy threats due to sequential background knowledge
attacks, and we compare our defense with other applicable 
solutions, in terms of both privacy protection and data 
quality.

\subsection{Experimental setup}
\begin{figure*}[t!]
\centering
\subfigure[l-div.]{\label{fig:semper-ldiv}
\includegraphics[width=.47\columnwidth]{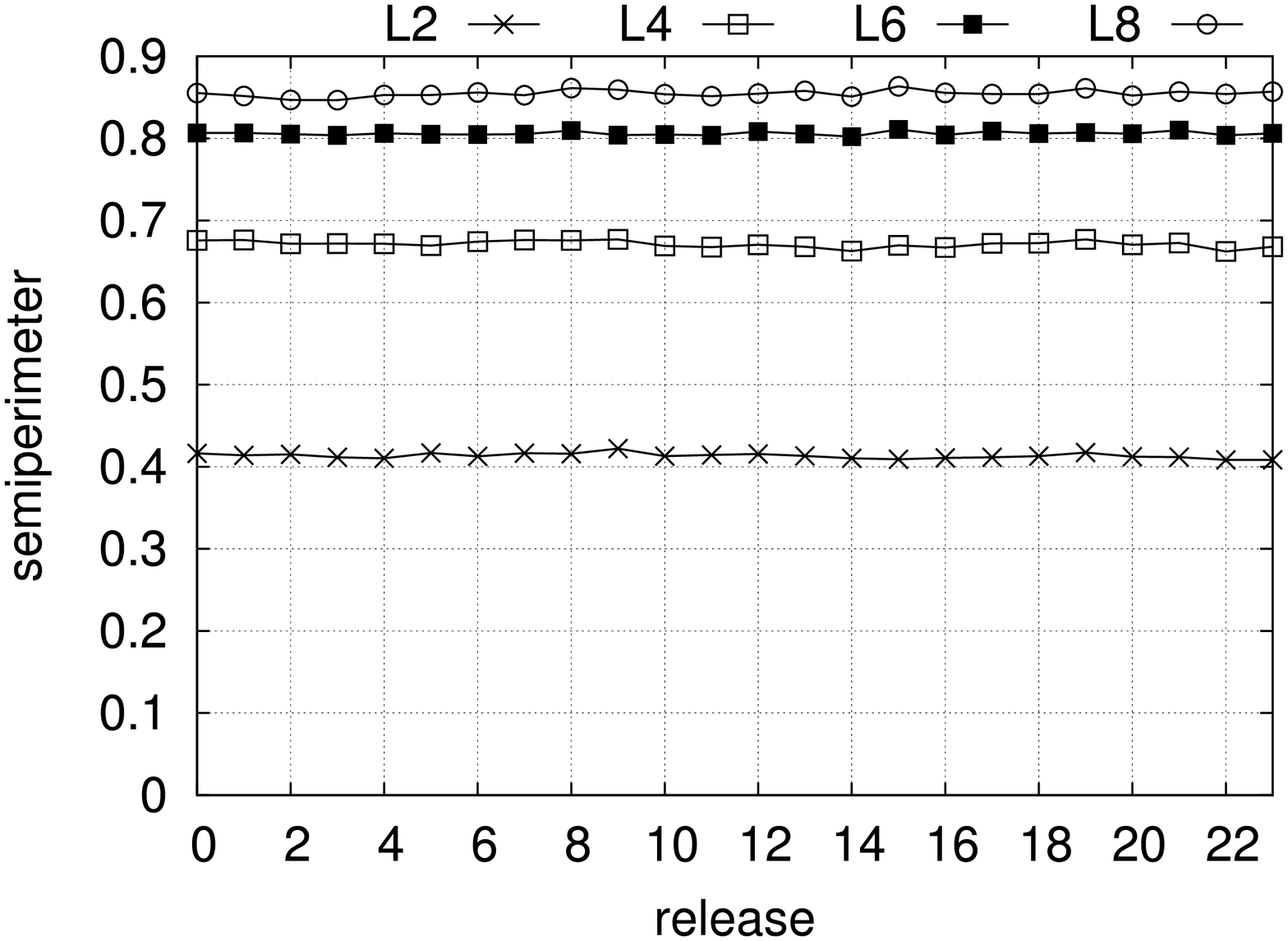}

}
 \subfigure[t-clos.]{\label{fig:semper-tclos}
 \includegraphics[width=.47\columnwidth]{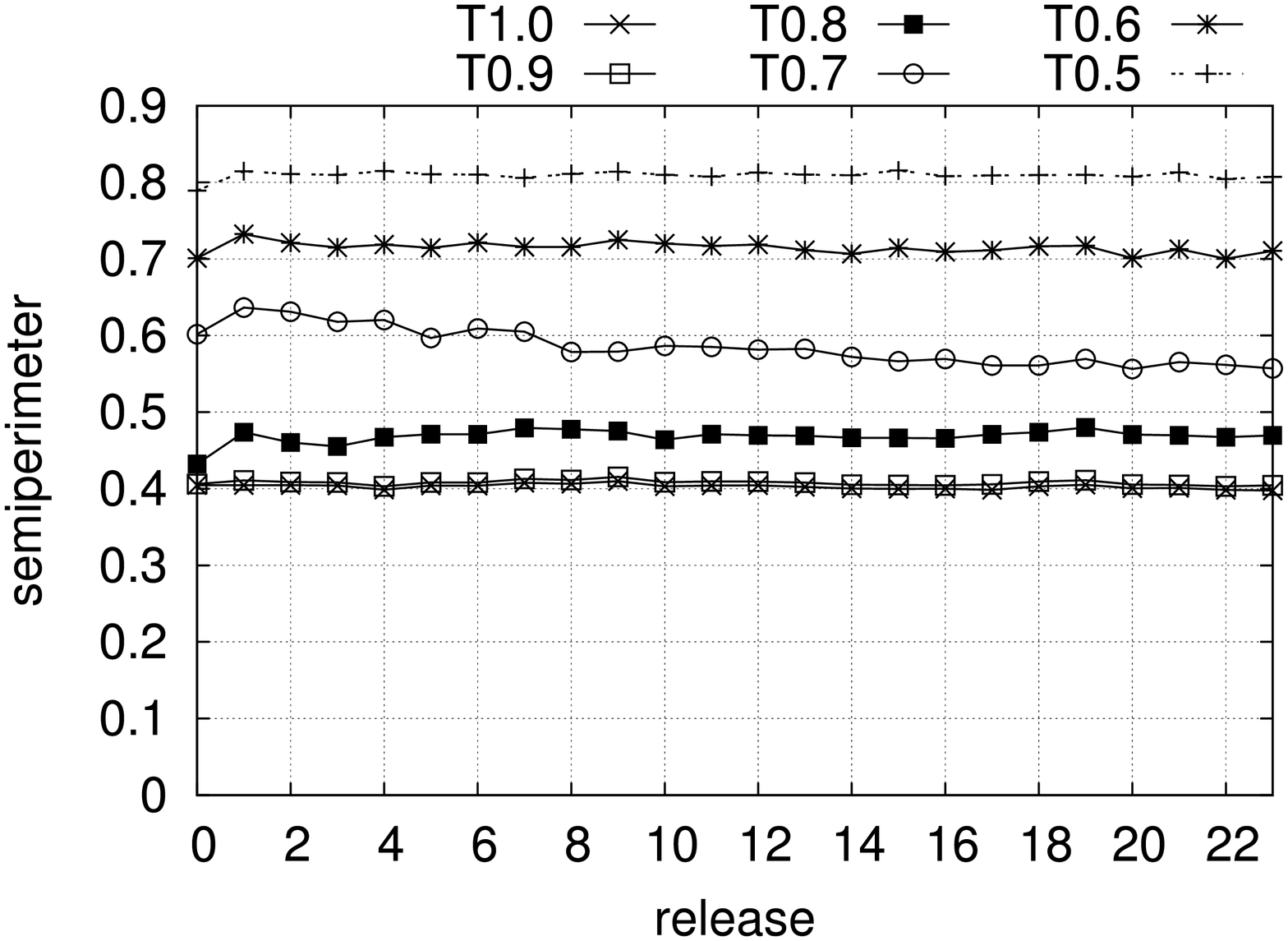}

}
 \subfigure[(B,t)-priv.]{ \label{fig:semper-bt}
 \includegraphics[width=.47\columnwidth]{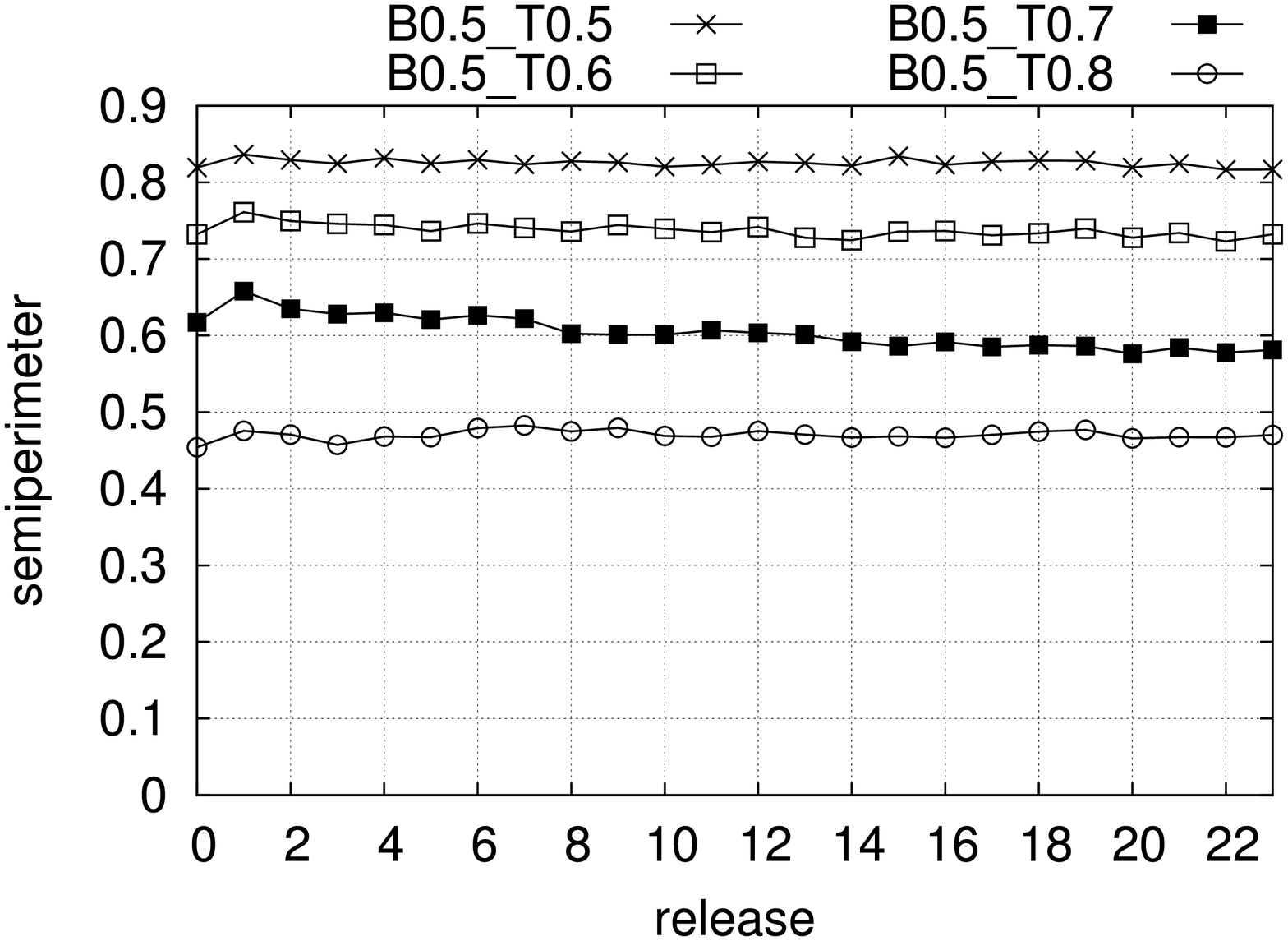}

}
 \subfigure[JS-red.]{ \label{fig:semper-tjs}
 \includegraphics[width=.47\columnwidth]{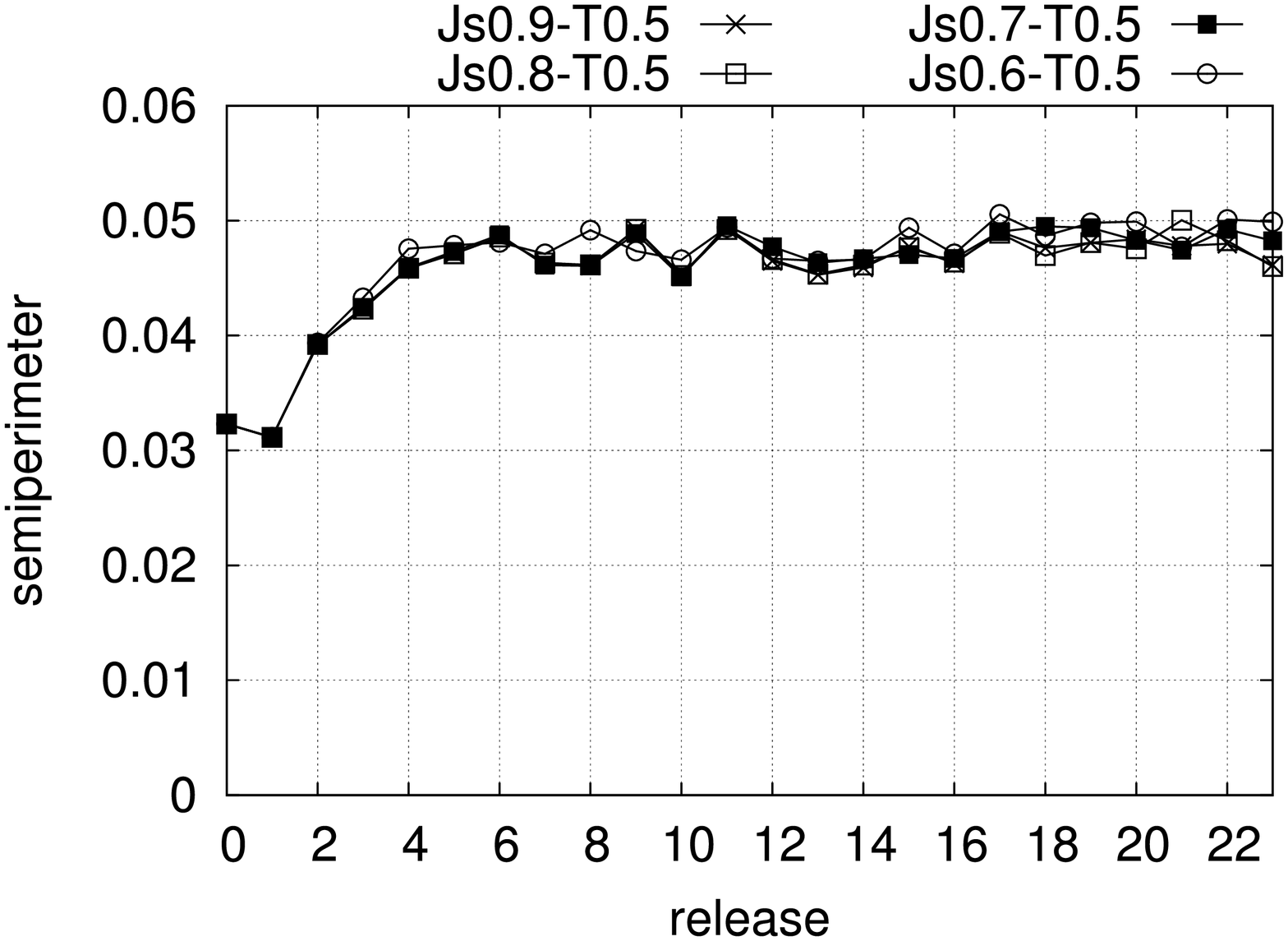}

}
\caption[]{\qi\ generalization}
\label{fig:semiperimeter}
\end{figure*}
\begin{table}[b!]
  \scriptsize
  \centering
  \begin{tabular}[h]{|c ||c|c|c|c|}
    \hline
      & l & t & B & j \\ \hline \hline
    l-div. &  $[2, 8] \, \mathbf{2}$  & - & - & - \\ \hline
    t-clos. & - & $[0.5, 1] \, \mathbf{0.8}$ & - & - \\ \hline
    (B,t)-priv. & - & $[0.5, 0.8] \, \mathbf{0.8}$ & $[0.3, 0.7] \, \mathbf{0.5}$ & - \\ \hline
    JS-red. & - & $[0.5, 0.8] \, \mathbf{0.5}$ & - & $[0.2, 0.8] \, \mathbf{0.6}$ \\ \hline
  \end{tabular}
  \caption{Privacy parameters used in the experiments}
  \label{tab:parameter}
\end{table}

To the best of our knowledge, all the datasets used for experimental
evaluation of proposed privacy defenses for serial data publication
were created from non-temporally characterized sets of tuples, in
which each tuple was randomly assigned to a release.  Clearly, these
datasets are not realistic for investigating the use that an adversary
can make of temporal correlations.  The dataset used in our
experiments has been synthetically created based on domain knowledge
extracted from the medical literature; in particular, studies reported
in~\cite{Sepsi,hiv,Hepatitis,Alzhaimer}. Each of those papers provides
the probabilities that a specific disease evolves from one stage to
another based on the characteristics of the patient (age, gender and
weight) and on the past evolution of the disease.  Based on that
information, we computed $\bkseq$ as the probability of a patient
performing an exam at $\tau_i$ to obtain a given result
\emph{ex-res}$_i$ given a sequence of %
results of exams performed by that person in the previous weeks.
$\bksv$ was calculated %
dividing age and weight into $3$ sub-intervals (each one
containing $10$ values), and assigning different probability
distributions to each of the $18$ classes of users obtained
combining age, weight and gender values.
The dataset has been made available from our group and can be used 
to replicate our experiments, or as a testbed for any research about 
sequential background 
knowledge\footnote{\url{http://webmind.dico.unimi.it/BKseq-dataset.zip}}.

Experiments were performed on a history of 24 views, each one
containing 5,000 tuples. A total of 16,160 individuals appear
in at least one view of the history.
Tuples in the dataset represent the results of medical exams
performed in a given institute. 
One view per week is released, and each view contains the records
of exams performed during that week. A tuple is composed of
$3$ \qi\
attributes \emph{age}, \emph{gender} and \emph{weight}, and
a sensitive attribute \emph{Ex-res}. \emph{Age} has
values in the interval $[45,\ 74]$, \emph{gender} in
$[1, \ 2]$, and \emph{weight} in $[60, 89]$. The domain of
\emph{Ex-res} includes $17$ different values associated to stages of
different diseases ($5$ stages of liver disease, $4$ of the HIV syndrome,
$3$ of Alzheimer, and $5$ of sepsis), as well as two sensitive
values to describe the \emph{deceased} and \emph{discharged} events.

Since our study is the first to
consider the role of sequential background
knowledge in privacy-preserving data publishing, a direct comparison
with techniques specifically devoted to protect against the identified 
threats was not possible. 
However, we performed experiments to compare \defense\ with
state of the art privacy protection methods that are applicable
to our case: \emph{a)} \dld\ (each \QIG\ must contain at least $l$ 
tuples having different sensitive values), \emph{b)} \tc~\cite{LiICDE07}, 
and \emph{c)} \bt~\cite{LiICDE09}.
We used the Mondrian framework~\cite{LeFevreICDE06} to generalize the 
views in the history according to each of the latter methods, while
we used Algorithm~\ref{alg:general-defense} to apply the \defense\
defense.
Experiments were performed on a 2.4GHz workstation with 4GB RAM. 
The time required for anonymizing a view with the \defense\ algorithm 
varied from a few minutes to a maximum of 43 minutes, depending on the 
chosen privacy parameters; this is an acceptable time since in many
cases anonymization is performed offline. 

For each considered technique, we made experiments with different 
values of the corresponding privacy parameters.
Figure~\ref{fig:semiperimeter} shows the average 
semiperimeter\footnote{The semiperimeter of a \QIG\ is the sum of 
the normalized lengths of the interval of each QI value of tuples 
in it.} of QI-groups generated by the different techniques using
the values shown in Table~\ref{tab:parameter}
(bold numbers indicate the parameters used in the following 
experiments). A smaller semiperimeter corresponds to a better 
quality of released data.

\SetAlFnt{\footnotesize}
\linesnumbered

\begin{algorithm}[t!]
\caption{$\spmbkseq$ extraction}
\label{alg:spm}

 \dontprintsemicolon
 \KwIn{History of original views $\mathcal{H}_r = \langle V_{1}, \ldots, V_{r} \rangle$, 
  a sequence of sensitive values $seq$, and a
   sensitive value $s$.}

 \KwOut{The conditional probability $p(s | seq )$,
 which corresponds to the frequency of sequence $\langle seq, s \rangle$ in
 $\mathcal{H}_r$.}

\BlankLine

 \textbf{SPM$(\mathcal{H}_r, seq, s)$}
 \Begin{
   \For{$h = 1$ to $r$}{
     \ForAll{respondent $u$ of a tuple in $V_h$}{
       \For{$j = h$ to $1$}{
         $seq_j$ = seq. of past $j$ sensitive values of $u$ in $\mathcal{H}_h$\;
         $seq_j.numOcc = seq_j.numOcc + 1$\;
       }
     }
   }
   \lIf{(seq.numOcc == 0)}
   	{\Return{$0$}\;}
   	\Else{$sequence$ = $\langle seq, s \rangle$\;
	   \Return{$\displaystyle{\frac{sequence.numOcc}{seq.numOcc}}$}}
 }
\end{algorithm}

\subsection{Measuring the adversary gain of knowledge}
In order to evaluate the privacy threat, %
we measured the \emph{gain of knowledge} when an
adversary is able to exploit 
sequential background knowledge.
For a given generalized view $V^*_i$ released at $\tau_i$
containing $N$ tuples, we measured the \emph{average adversary gain}
$\rho$ as follows:
\[
   \rho = \frac{1}{N} \sum_{j = 1}^{N}\Bigg(\frac{p(r_j,s_{i_j}, \tau_i) - 
   \frac{m(s_{i_j})}{|Q_{i_j}|}}{1 - \frac{m(s_{i_j})}{|Q_{i_j}|}}\Bigg),
\]
where: $p(r_j,s_{i_j}, \tau_i)$ is the value of posterior knowledge
computed based on background knowledge
for respondent $r_j$ and her actual 
private value $s_{i_j}$ at $\tau_i$;
$Q_{i_j}$ is the \QIG\ of $V_i^*$ containing
the tuple whose respondent is $r_j$; and $m(s_{i_j})$ is the number of
tuples $t$ in $Q_{i_j}$ such that $t[S] = s_{i_j}$. 
Intuitively, the adversary gain represents the 
amount of information
obtained with the use of %
background knowledge with respect to a privacy attack 
based only on the observation of the frequency of sensitive values
in the \QIG.

\begin{figure*}[t!]
\centering
 \subfigure[l-div.]{\label{fig:gainVsKnow-ldiv}
 \includegraphics[width=.64\columnwidth]{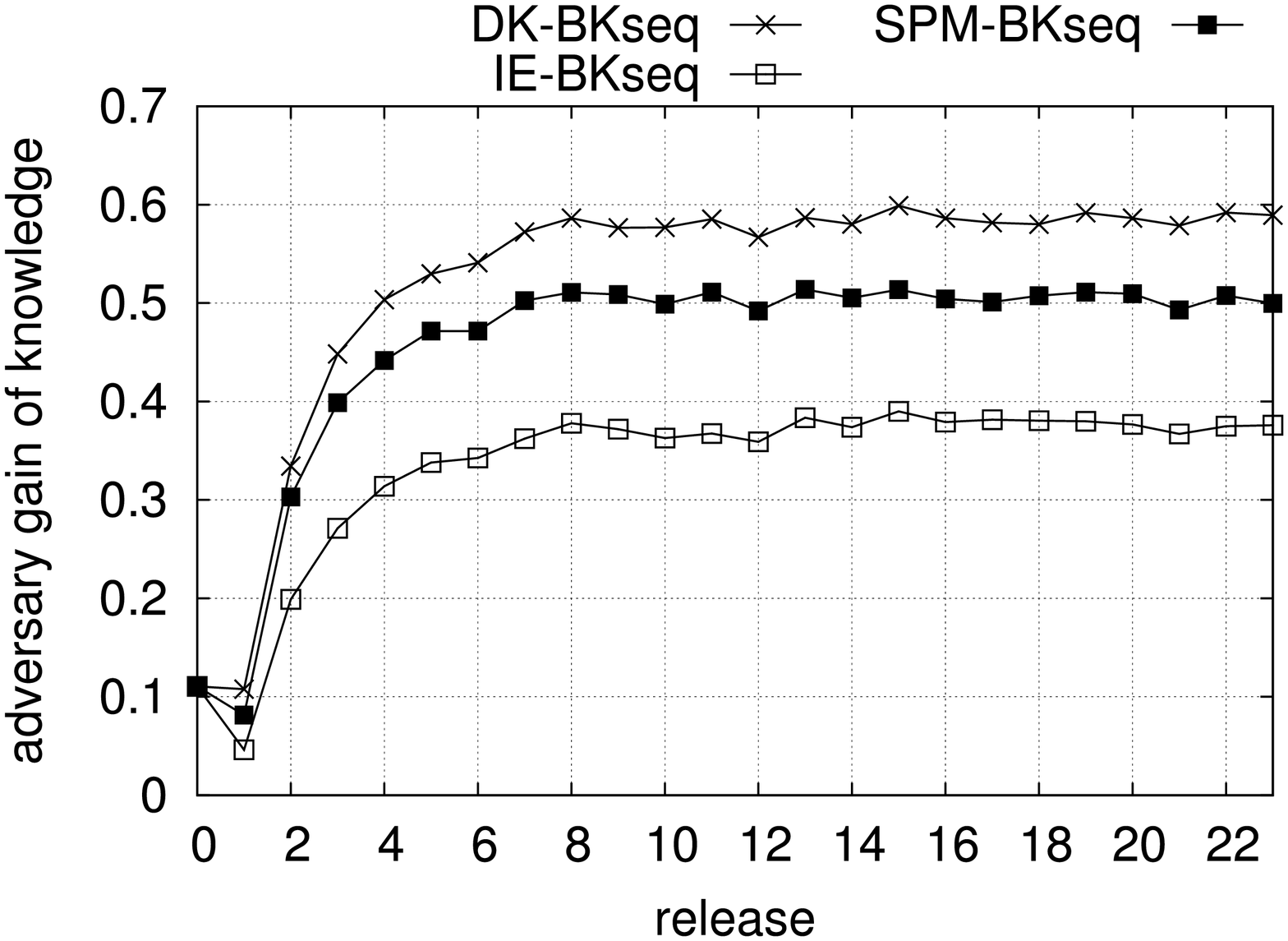}
}
 \subfigure[t-clos.]{\label{fig:gainVsKnow-tcl}
 \includegraphics[width=.64\columnwidth]{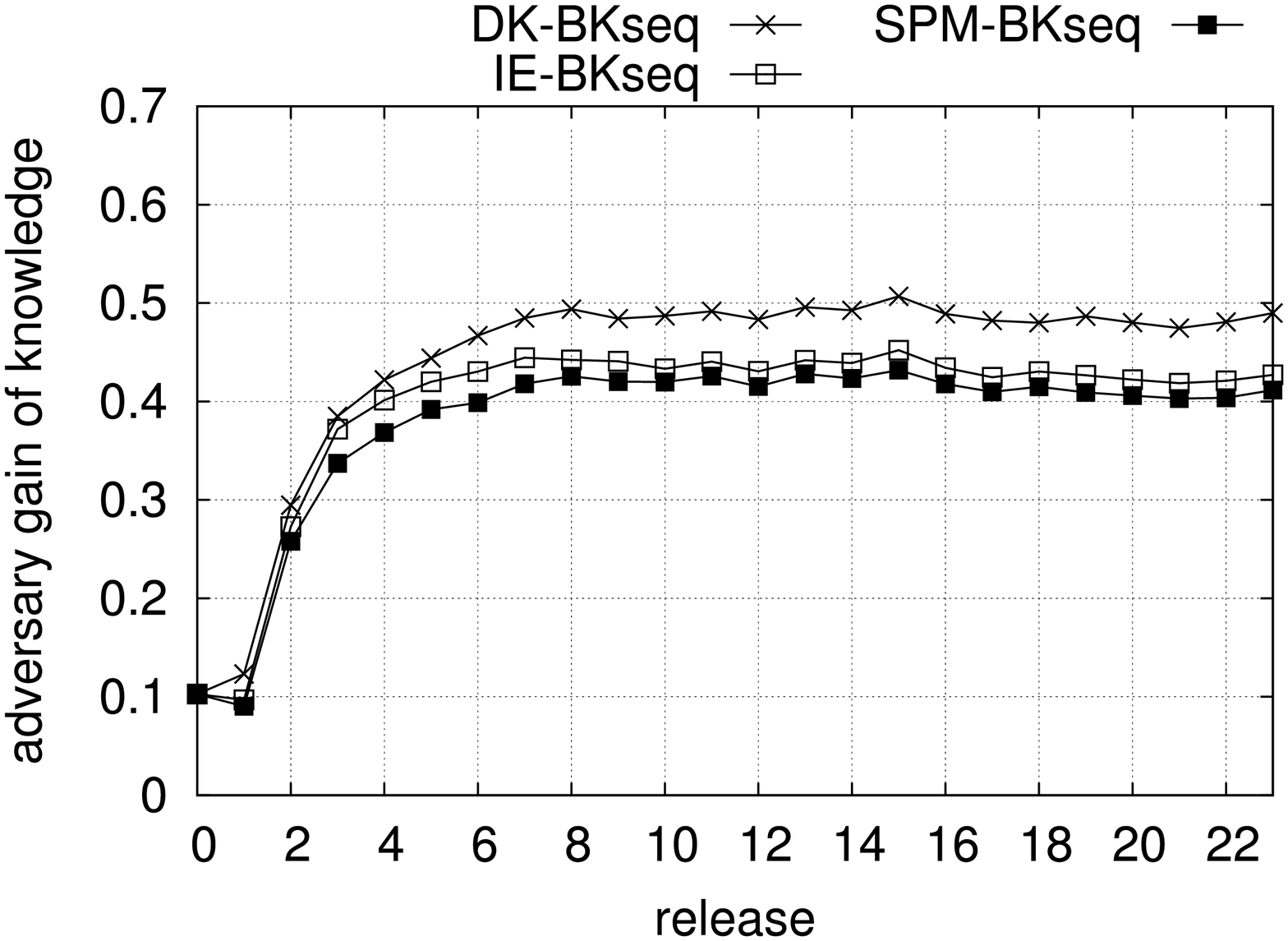}
}
\subfigure[(B,t)-priv.]{\label{fig:gainVsKnow-bt}
 \includegraphics[width=.64\columnwidth]{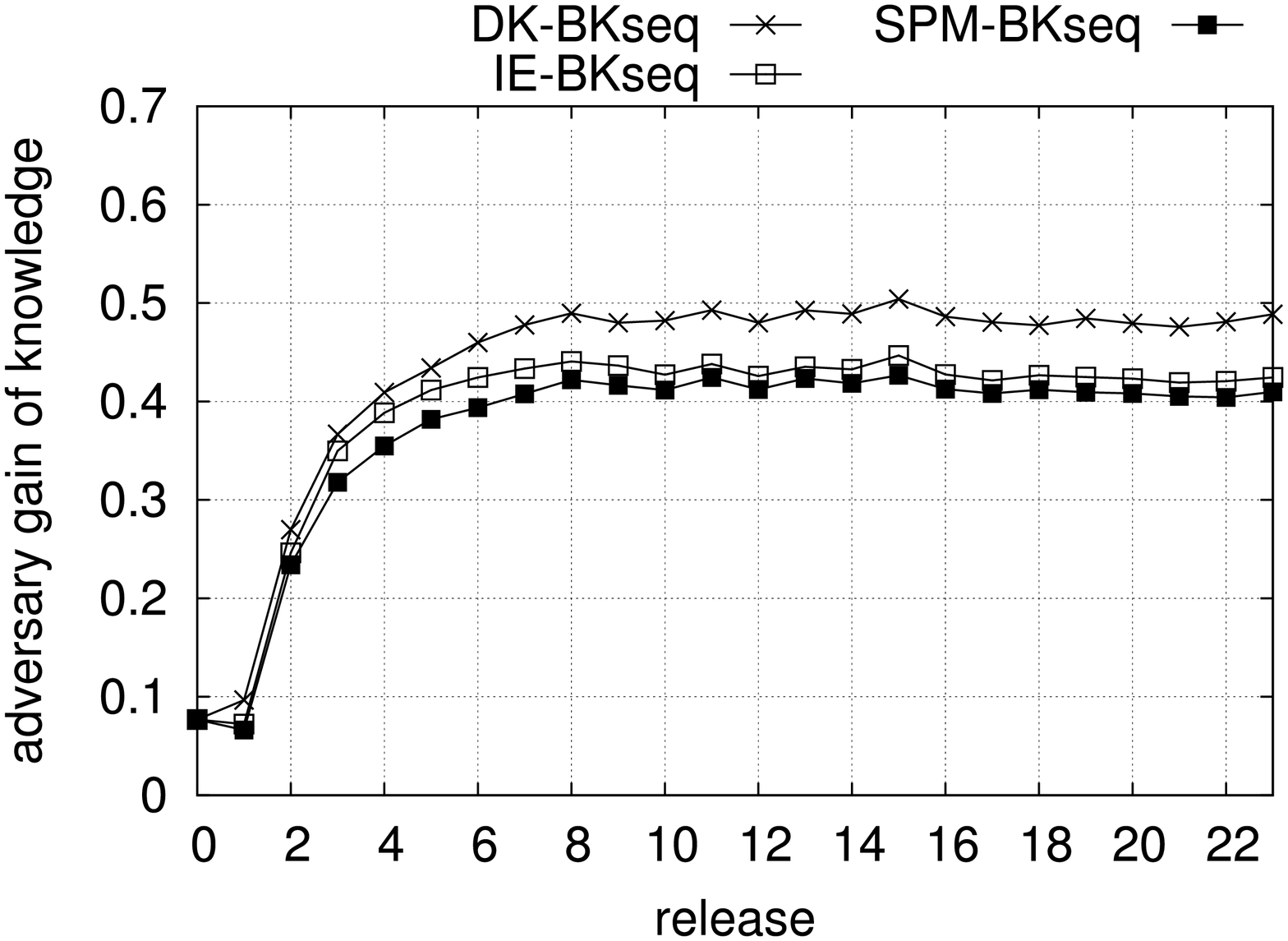}
}
\caption[]{Adversary gain vs different kinds of adversary's $\bkseq$}
\label{fig:gainVsKnow}
\end{figure*}
\begin{figure*}[t!]
\centering
 \subfigure[l-div.]{\label{fig:gain-ldiv}
 \includegraphics[width=.64\columnwidth]{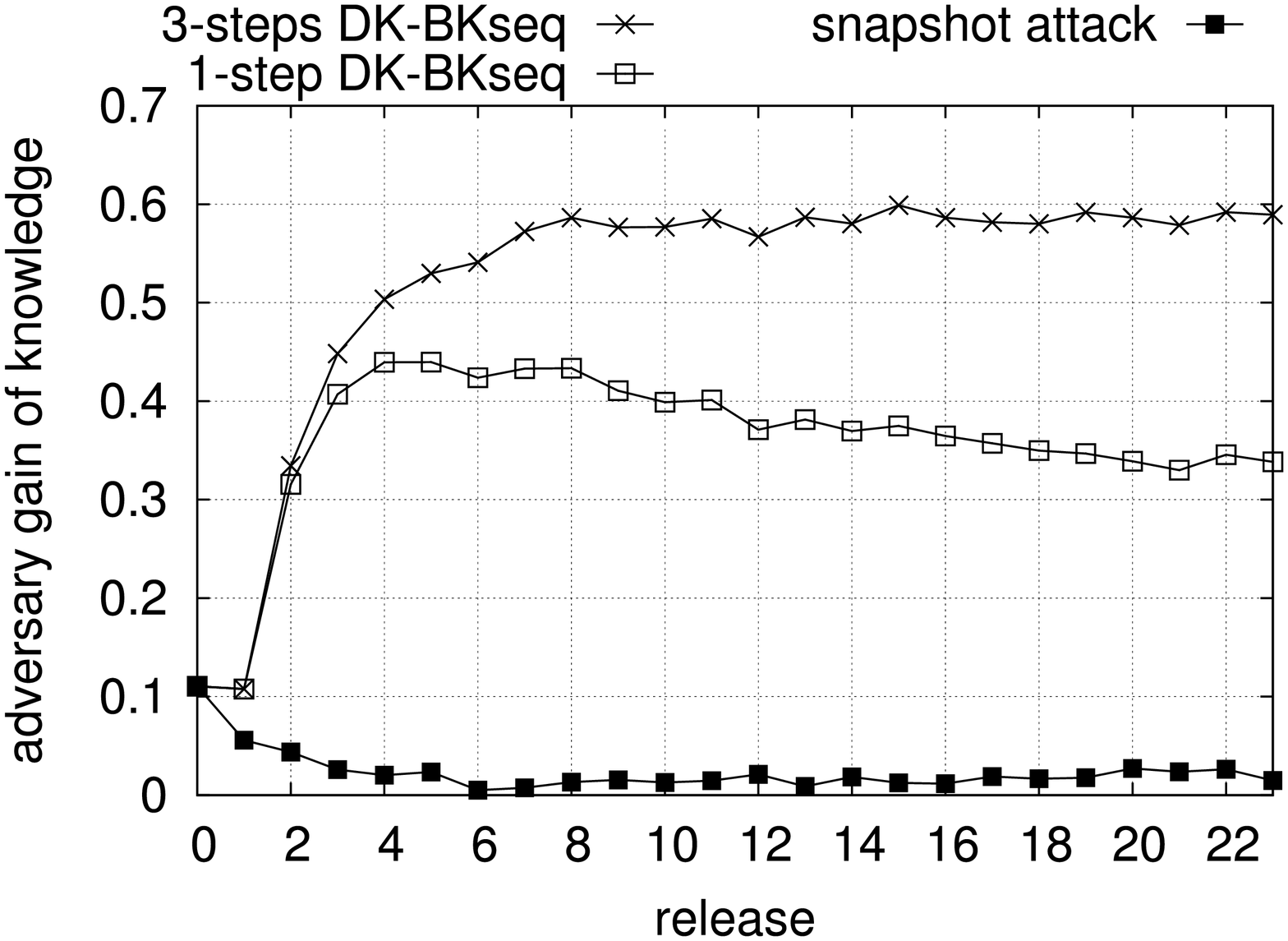}
}
 \subfigure[(B,t)-priv.]{ \label{fig:gain-bt}
 \includegraphics[width=.64\columnwidth]{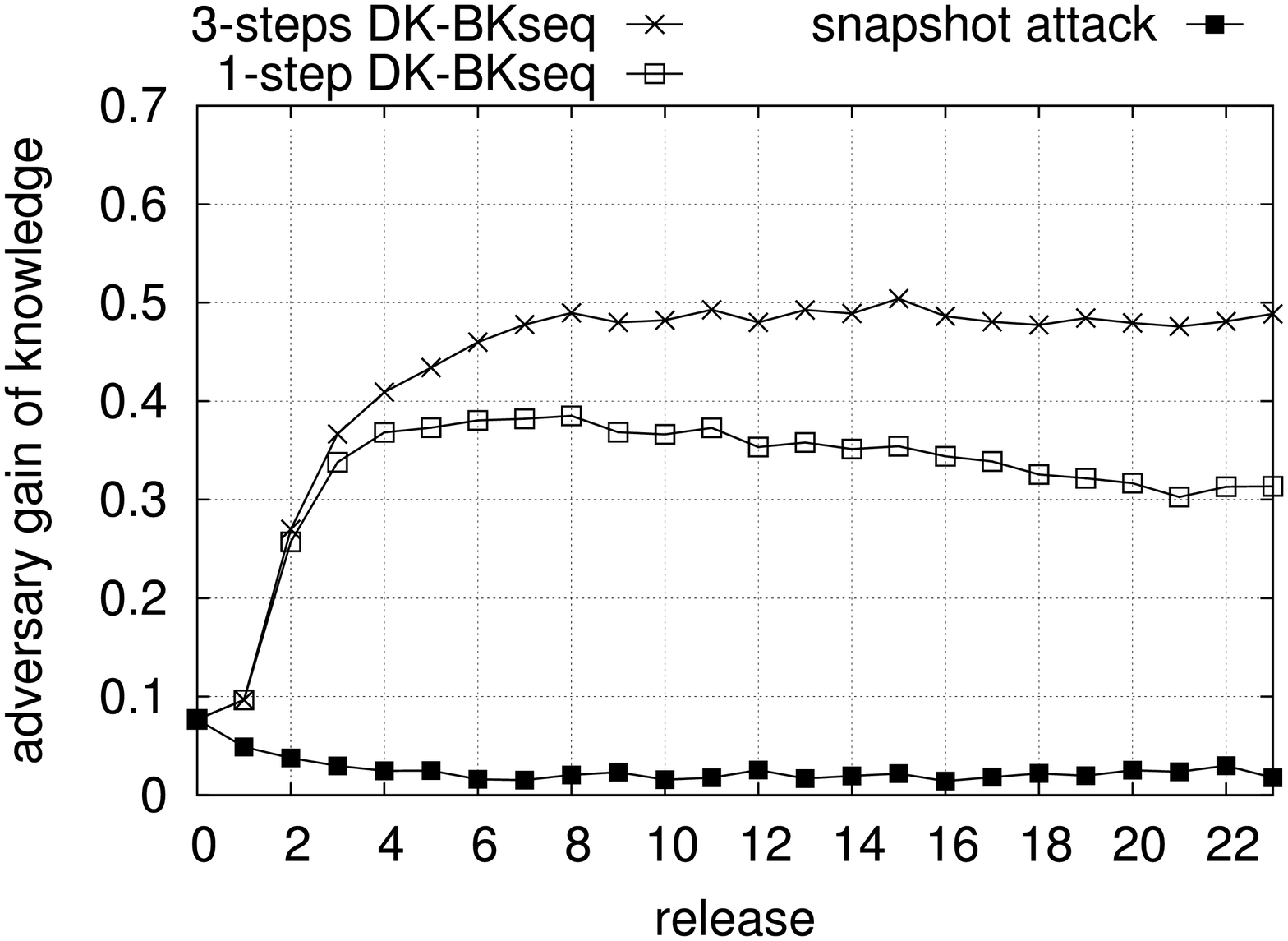}
}
 \subfigure[JS-red.]{ \label{fig:gain-tjs}
 \includegraphics[width=.64\columnwidth]{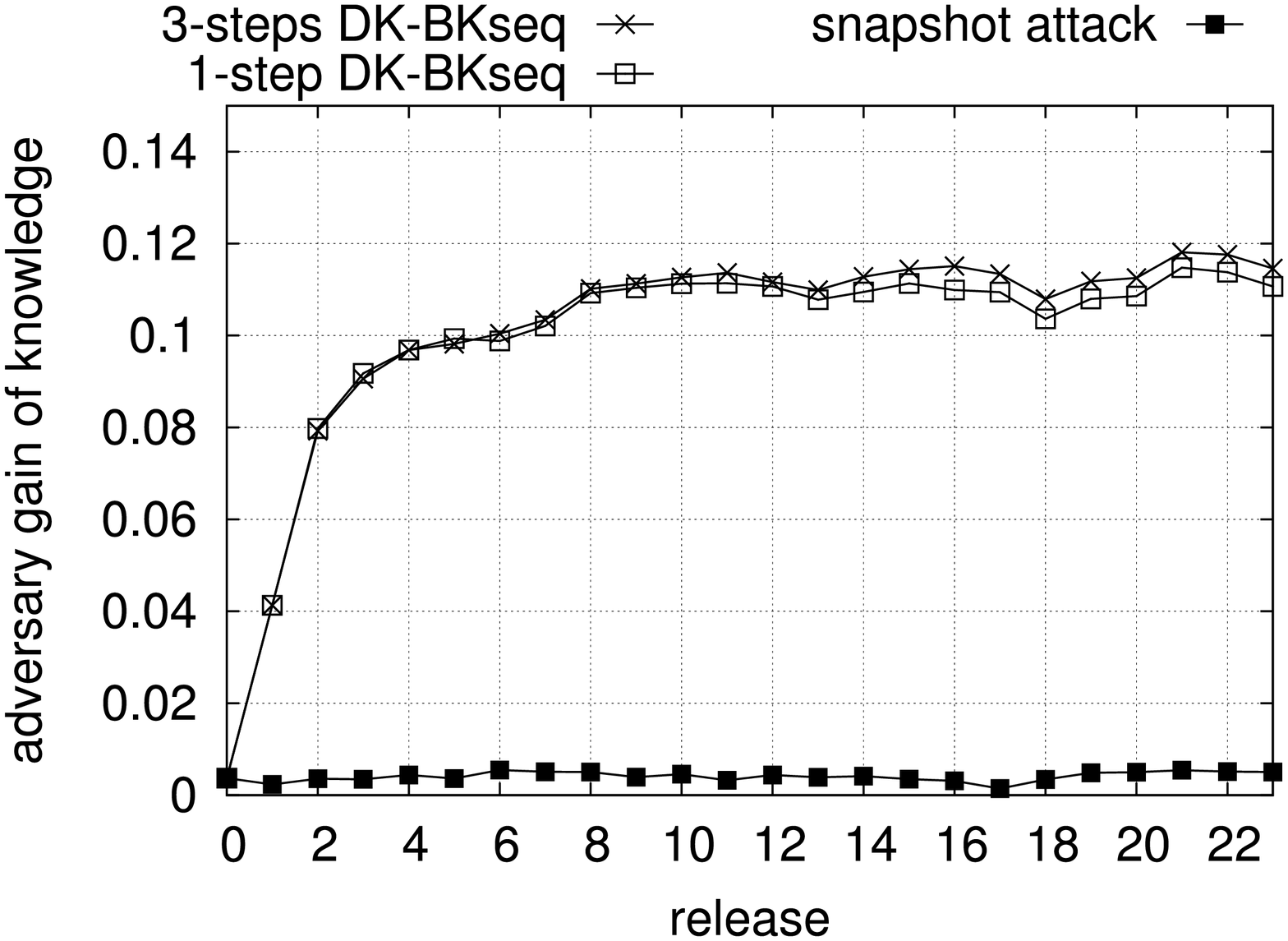}
}
\caption[]{Adversary gain vs accuracy of adversary's domain knowledge $\dkbkseq$}
\label{fig:gain-app}
\end{figure*}

\subsection{The role of adversary's background knowledge}
\label{subsec:exp-bkext}
We performed experiments to evaluate the role of background
knowledge on the privacy threats investigated in this paper:
\squishlist
\item \emph{Incrementally extracted knowledge} $\iebkseq$. 
Since it was the subject of related studies (e.g.,~\cite{Li08,LiICDE09}),
the first kind of background knowledge we consider is 
the one
directly extracted from the data to be released.
$\iebkseq$ can be calculated by applying sequential pattern mining (SPM) 
techniques on the history of original (i.e., non-anonymized) data;
at each time $\tau_i$, $\iebkseq$ is calculated based on $V_i$.
Since the size of the corpus is relatively small, we applied a simple SPM
algorithm, which is essentially based on a frequency count of sequences 
appearing in the history. The algorithm is illustrated in
Algorithm~\ref{alg:spm}.
\item \emph{Mined knowledge} $\spmbkseq$. In practice, an adversary may
approximate $\bkseq$ by applying SPM techniques
on an external corpus of non-anonymized data. We created a data corpus using the 
same model that we used to generate our dataset; the corpus consists in 
a history of 24 views containing 5,000 tuples each.
$\spmbkseq$ was calculated by applying Algorithm~\ref{alg:spm} to that
corpus.
\item \emph{Domain knowledge} $\dkbkseq$. 
Since the dataset we used was generated based on domain knowledge, 
in our experiments $\dkbkseq$ corresponds to the exact $\bkseq$;
i.e., it is the ``best'' knowledge that an 
adversary may have.
However, in general 
an adversary's domain knowledge may only approximate the exact $\bkseq$.
Hence, we also considered another kind of domain knowledge,
whose temporal extent is limited to a
number $n$ of past observations.
We denote this knowledge as $n$-\emph{steps} $\dkbkseq$, and we consider
$n=1$, $n=2$, and $n=3$.
\squishend

Figure~\ref{fig:gainVsKnow} shows the adversary gain when views are
anonymized using existing techniques, and the adversary may exploit 
the different kinds of sequential background knowledge. Results show 
that existing techniques are 
not effective against the attacks identified in this paper. Indeed, 
with each kind of background knowledge, the adversary gain grows very 
rapidly during the first 6/8 releases, exceeding the value of $0.4$.

For each considered anonymization technique, the form of
background knowledge that determines the highest adversary gain is full
$\dkbkseq$, since in our experiments it corresponds to the exact $\bkseq$.
Hence,  we considered
approximate $\dkbkseq$ in order to better evaluate the role of domain knowledge. 
Results illustrated in Figures~\ref{fig:gain-ldiv} and~\ref{fig:gain-bt}
show that even attacks based on approximate $\dkbkseq$
are effective against existing anonymization techniques;
attacks exploiting 
$3$-steps $\dkbkseq$ are more successful
than the ones exploiting $2$-steps and $1$-step knowledge
(we omit the plot for \tc\ since it is analogous to
the one for \bt).
Results also show that when the adversary exploits only $\bksv$ (i.e.,
when he performs a \emph{snapshot} attack), the gain of
information
with
respect to an attack considering only the frequency of sensitive values
is negligible.
The descending shape of curves for the $1$-step and snapshot attacks
is due to the fact that the background knowledge used by the adversary
tends to diverge from the one that generated the data, having a 
different temporal characterization.

\begin{figure*}[t!]
\centering
 \subfigure[Defense based on $\dkbkseq$]{ \label{fig:anonVsKnow-DK}
 \includegraphics[width=.64\columnwidth]{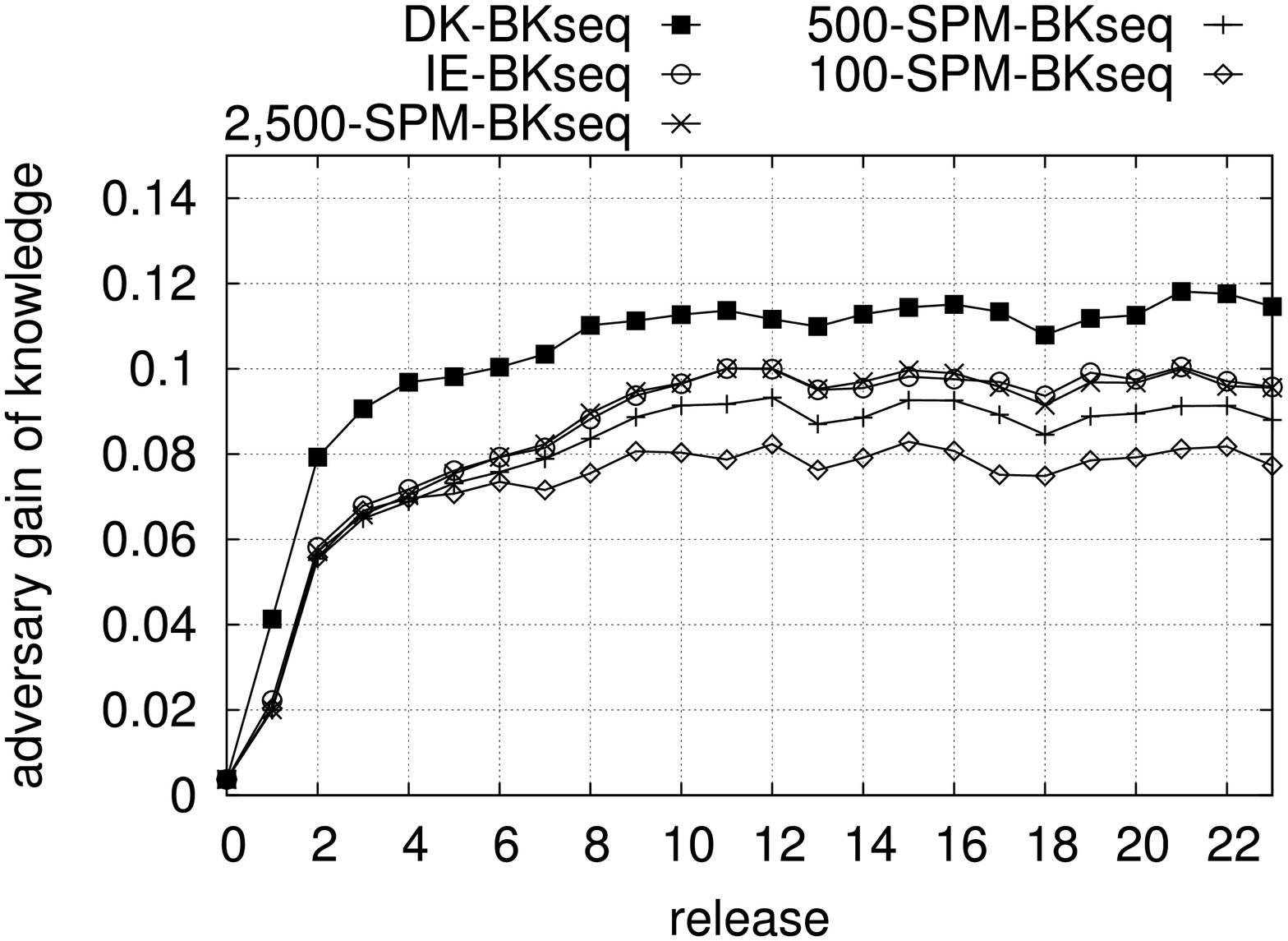}
}
\subfigure[Defense based on $\iebkseq$]{ \label{fig:anonVsKnow-IE}
 \includegraphics[width=.64\columnwidth]{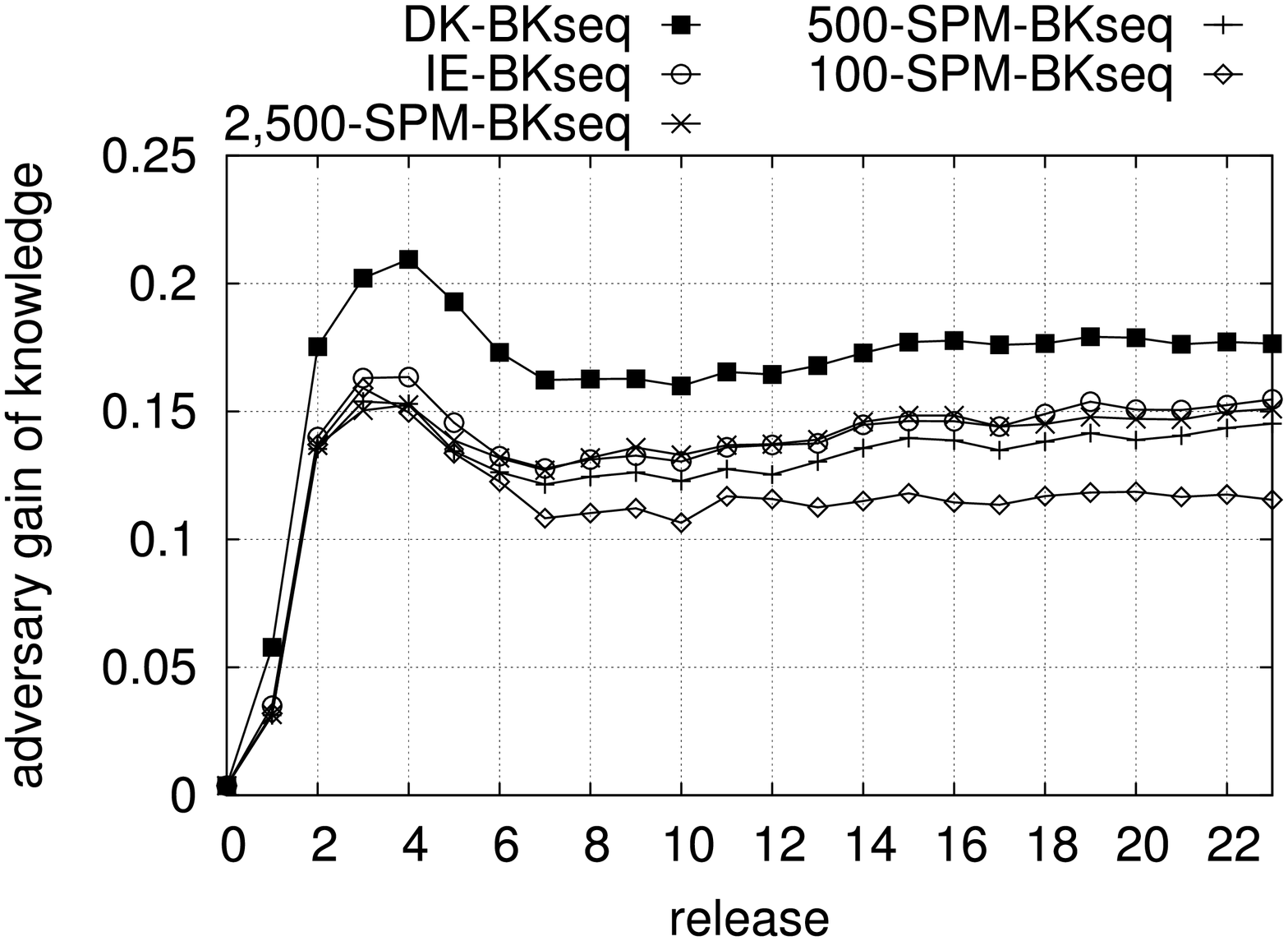}
}
\subfigure[Defense based on $\spmbkseq$]{ \label{fig:anonVsKnow-SPM}
 \includegraphics[width=.64\columnwidth]{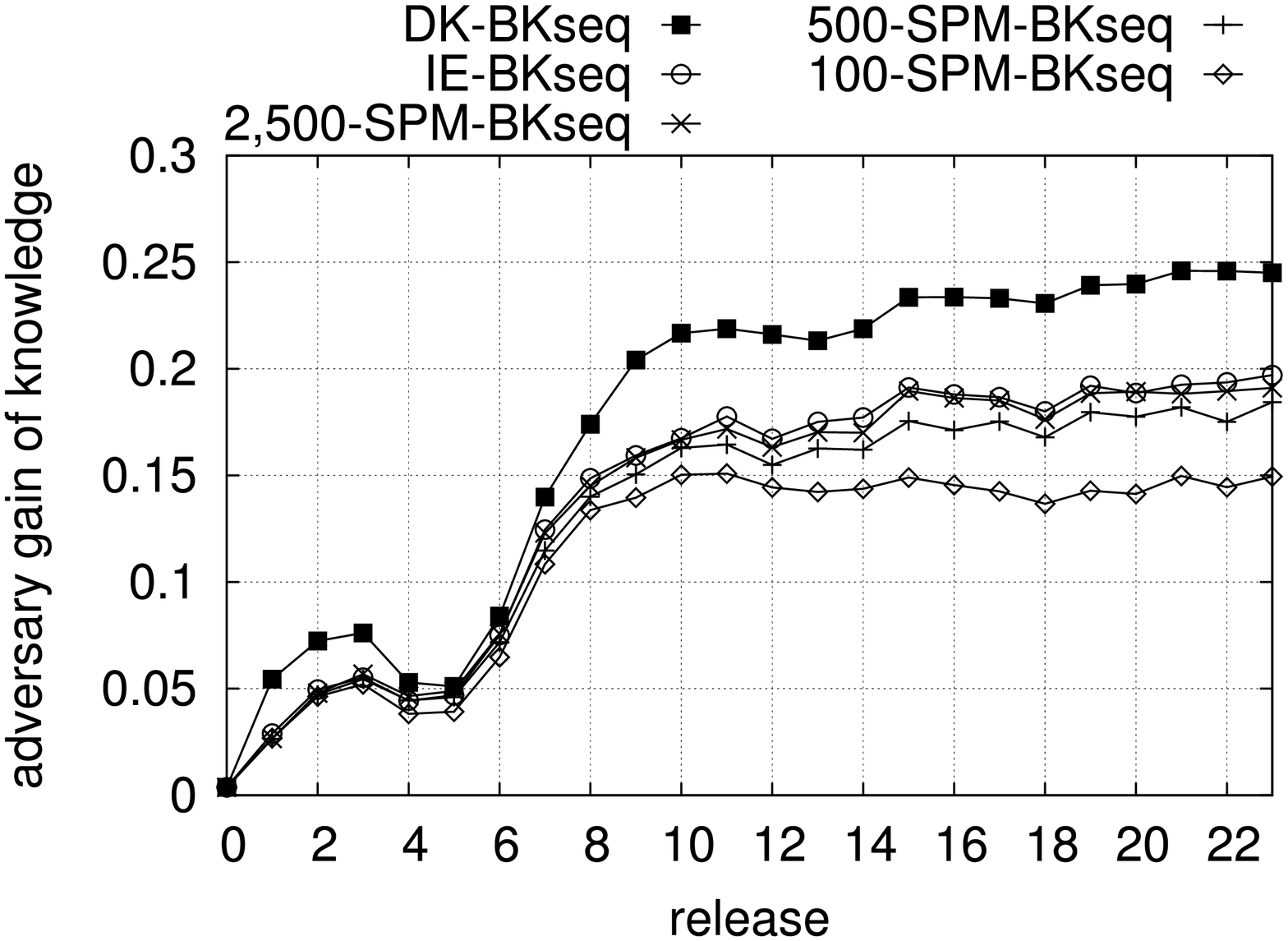}
}
\caption[]{\defense\ vs different kinds of adversary's $\bkseq$}
\label{fig:jsreduceVsKnow}
\end{figure*}
\begin{figure*}
\begin{minipage}[c]{0.65\columnwidth}
\centering
\includegraphics[width=1\columnwidth]{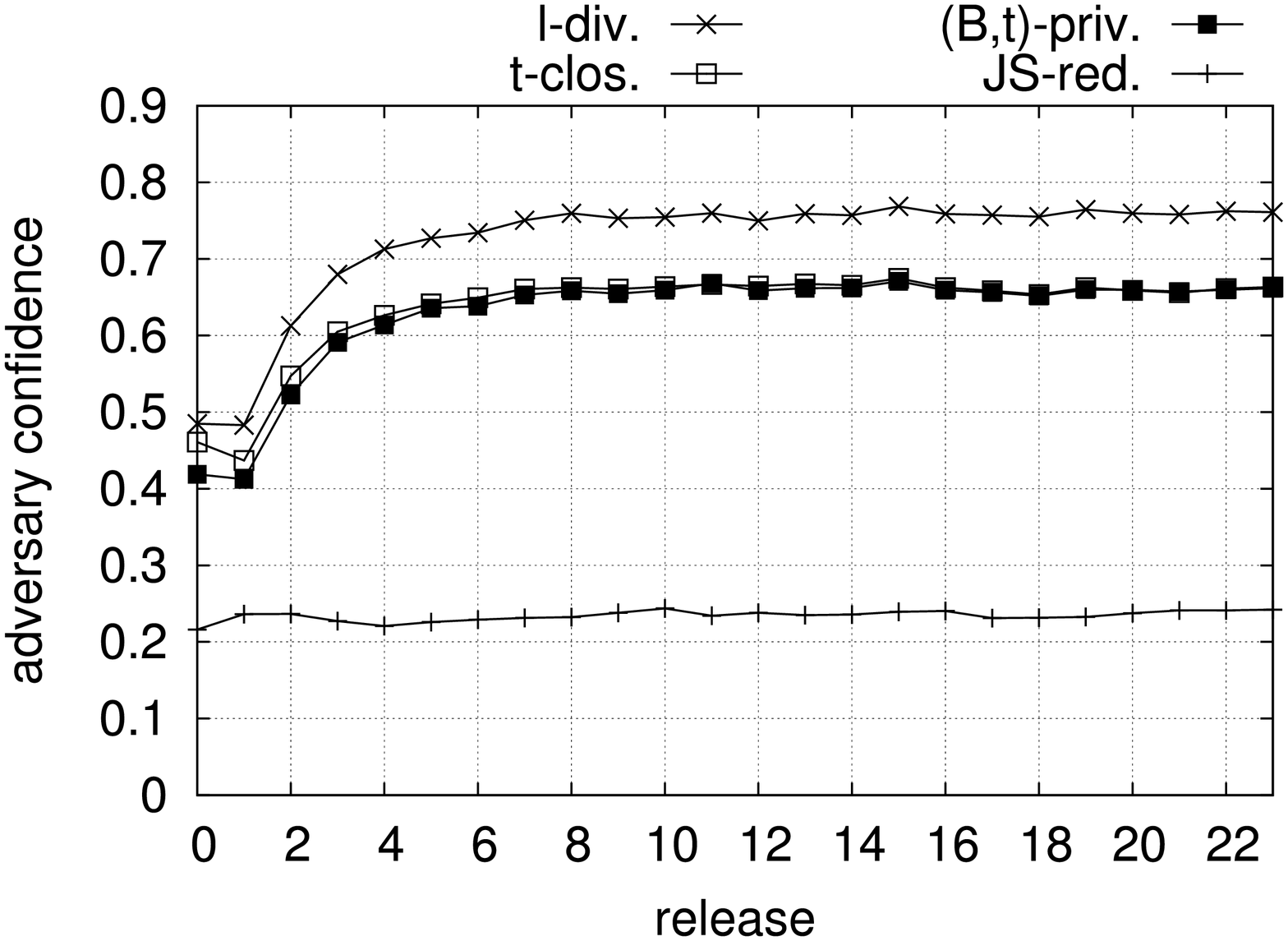}
\caption{Adversary confidence\label{fig:conf-app}}
\end{minipage}
 \hspace{1mm} 
\begin{minipage}[c]{1.4\columnwidth}
\centering
 \subfigure[GCP]{ \label{fig:gcp-app}
 \includegraphics[width=.45\columnwidth]{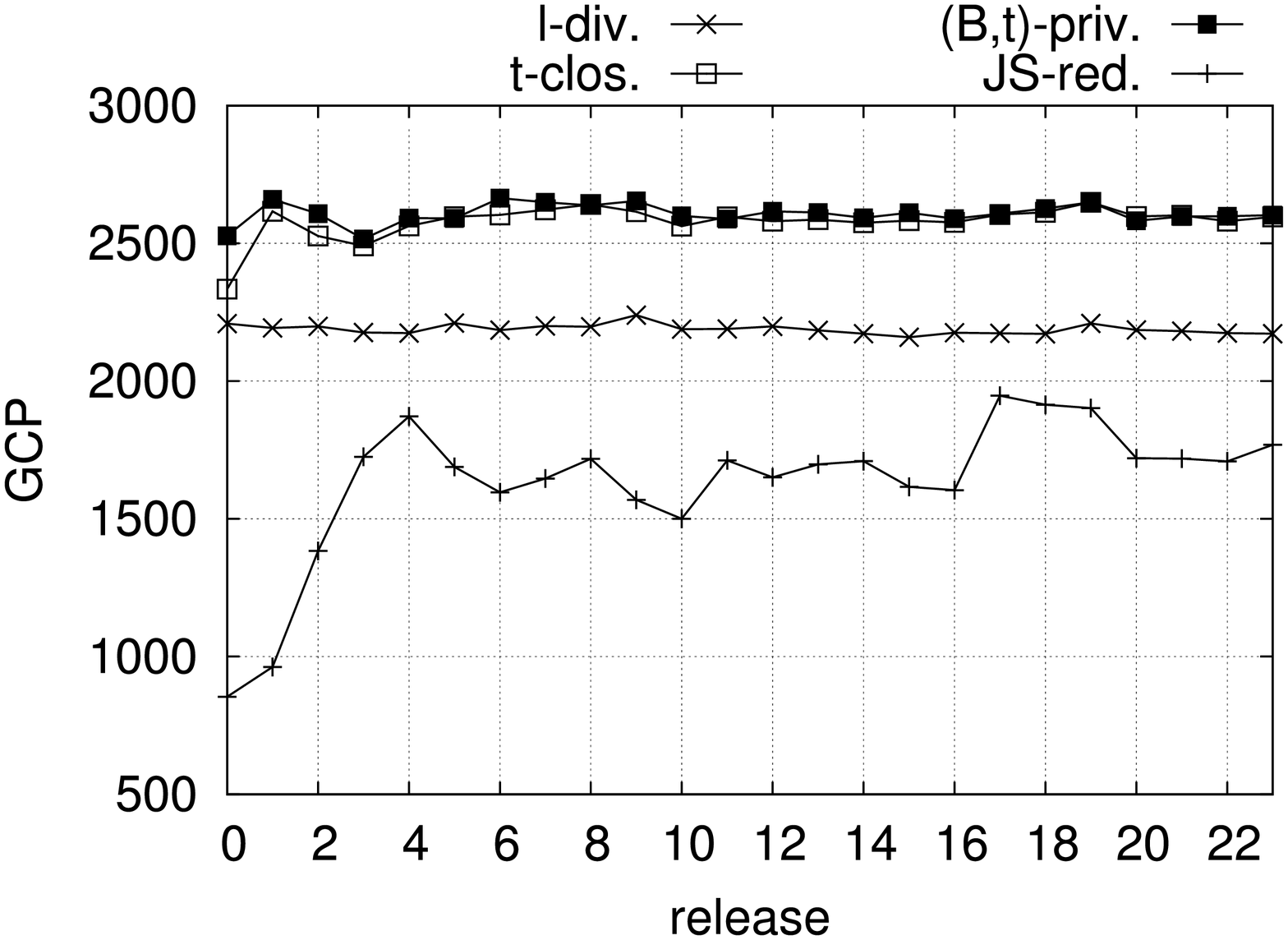}

}
\subfigure[Query error]{\label{fig:queryError}
 \includegraphics[width=.45\columnwidth]{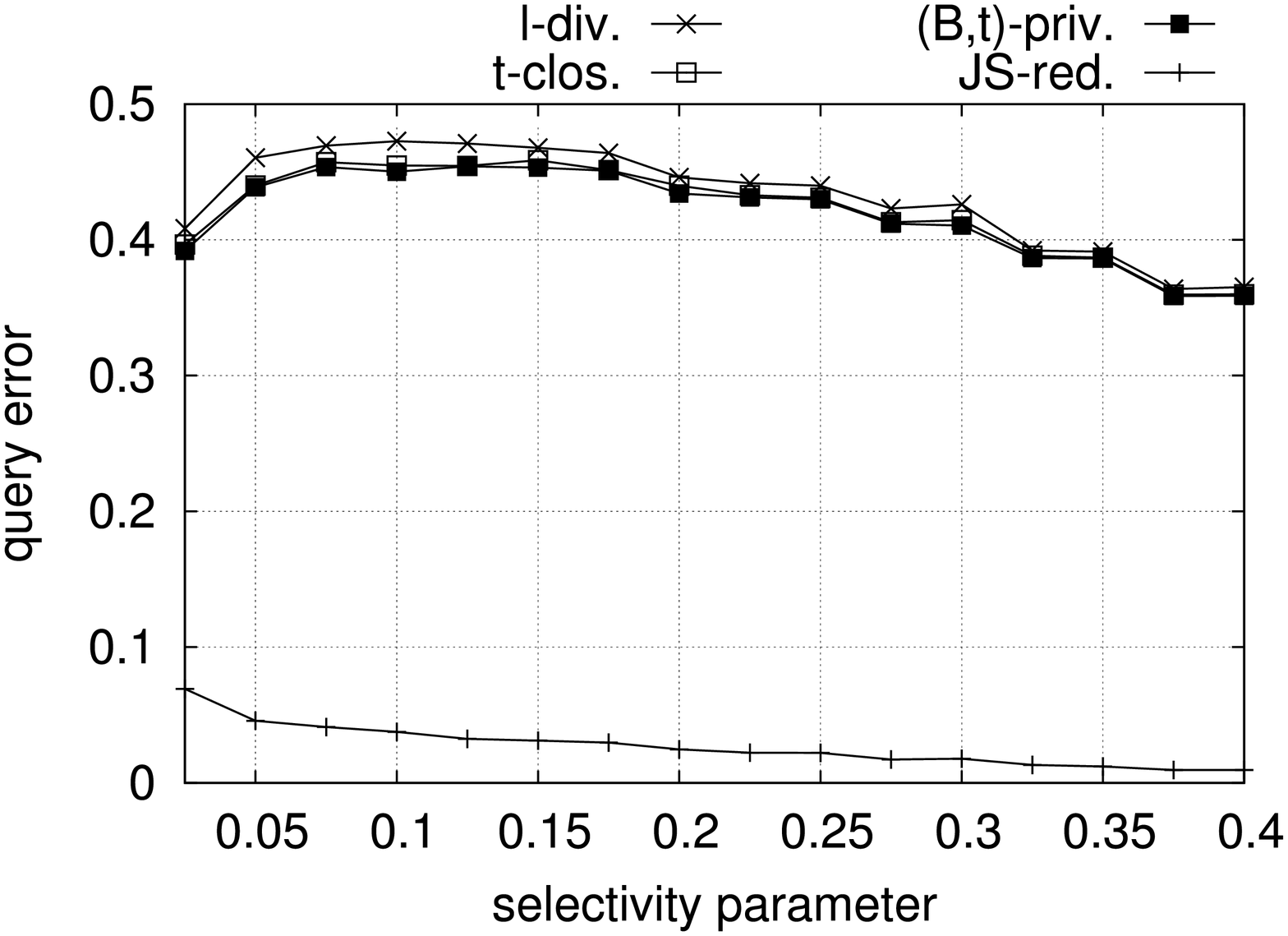}

}
\caption[]{Data quality evaluation}
\end{minipage}
\end{figure*}

\subsection{Effectiveness of the JS-reduce defense}
\label{subsec:effectiveness}
Experimental results reported in Figure~\ref{fig:gain-tjs} show that, 
when views are anonymized with the \defense\ technique, %
the adversary 
gain remains below $0.12$, independently from the length of the released 
history, and on the kind of domain knowledge available to the
adversary. This result shows that \defense\ significantly limits the inference
capabilities of the adversary with respect to the other techniques
that lead to an adversary gain higher than $0.5$.

We performed other experiments to evaluate the effectiveness of
\defense\ with different combinations of background knowledge
available to the defender and to the adversary, respectively.
In Figure~\ref{fig:anonVsKnow-DK}, we considered the case in which
the defender has background knowledge $\dkbkseq$.
In this case, the defense is very effective, %
even when the adversary has the same background
knowledge as the defender. 
When the adversary's background knowledge is extracted from the data,
we observe that the adversary gain is lower. 
With the label $n$-$\spmbkseq$ in Figure~\ref{fig:jsreduceVsKnow}, we denote
that the adversary's $\spmbkseq$ is extracted based on a history of $24$
views containing $n$ tuples each. The adversary gain
is lower with smaller values of $n$, since the resulting $\spmbkseq$
is a coarser approximation of the exact $\bkseq$. The adversary
gain with incrementally extracted knowledge is comparable to the one
obtained with $\spmbkseq$.

We also considered the unfortunate case in which the adversary has more accurate
background knowledge than the defender. 
Results illustrated in Figures~\ref{fig:anonVsKnow-IE} and~\ref{fig:anonVsKnow-SPM} 
show the adversary gain when the defender's background knowledge
is $\iebkseq$ and $\spmbkseq$, respectively. As expected, the more
accurate the attacker's background knowledge with respect to the defender's
one, the more effective the attack. However, results show that
\defense\ provides sensible privacy protection even in the
worst case; indeed, the adversary gain always remains below $0.25$.
It is important to note that \defense\ is effective even when the 
defender has neither domain knowledge, nor external data to derive 
background knowledge. %
Indeed, even
extracting background knowledge from the data to be released, the
adversary gain is low.

In order to study in more detail the effectiveness of \defense, 
we considered a further metric, named 
\emph{average adversary confidence}.
We call \emph{adversary confidence regarding respondent $r$ at release
$\tau_j$} the value of the posterior probability
$\pksv(r,\tau_j)$ computed by the adversary for the actual private
value of $r$ at $\tau_j$. %
The average adversary confidence about a generalized view $V_j^*$ is the
average of the adversary confidence regarding respondents of tuples in
$V_j^*$.  Figure~\ref{fig:conf-app} shows a comparison among the considered privacy
techniques in terms of the adversary confidence with respect to the number
of observed anonymized views (attack and defense are based on $\dkbkseq$). 
These results show that with our technique
the adversary confidence does not significantly grow with respect to the
length of the release history. On the contrary, with the other techniques,
after a few anonymized views have been released, the 
adversary can %
predict with
high confidence the exact sensitive values of tuples
respondents.

\begin{figure}[t!]
\centering
\includegraphics[width=.85\columnwidth]{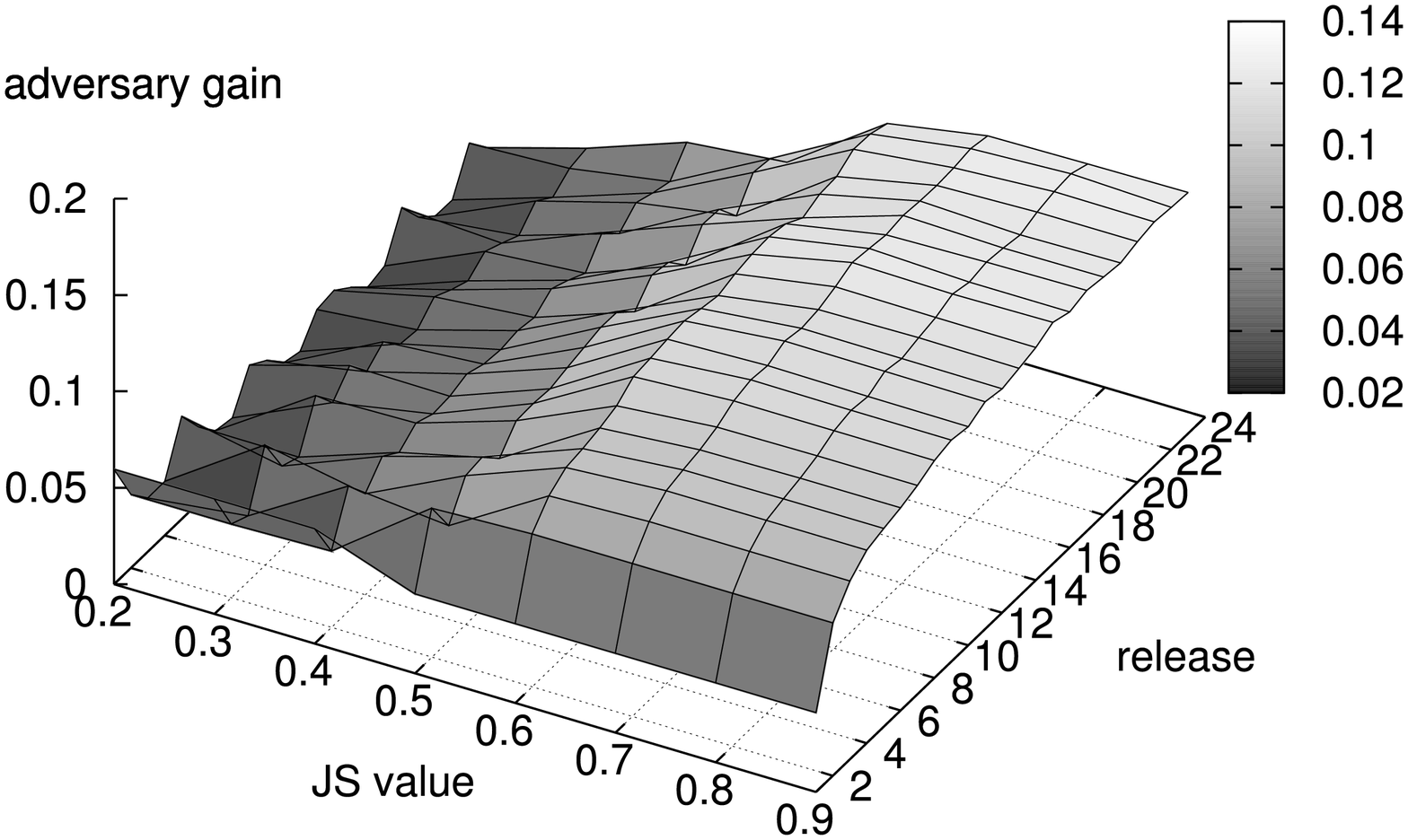}
\caption{Adversary gain versus \js\ ($t=0.5$)}
\label{fig:js-eval}
\end{figure}

We also performed specific experiments to evaluate the impact on
privacy protection of the \js\ threshold for the \defense\ defense.  
Results are illustrated in Figure~\ref{fig:js-eval}; as expected, the 
lower the JS threshold value, the lower the adversary gain.

\subsection{Data utility}
\label{subsec:utility}
In order to evaluate data utility, we considered both general utility
measures, and accuracy of aggregate query answering.
General utility is evaluated in terms of two well-known metrics:
average semiperimeter, and 
\emph{Global Certainty Penalty} (GCP)~\cite{XuKDD06} (a metric taking
into account the level of generalization of \qi\ values).
Figure~\ref{fig:semiperimeter}
shows the average semiperimeter of QI-groups generated by the
considered techniques (\defense\ is based on $\dkbkseq$). 
As it can be seen, \defense\ outperforms the
other techniques. %
These results are confirmed by a comparison in terms of GCP 
(Figure~\ref{fig:gcp-app}).

Then, we compared the utility of transaction data generalized by the different 
techniques in terms of the precision in answering aggregate queries (e.g., 
\emph{``count the number of individuals in the table whose QI-values belong to 
certain ranges"}). Queries were randomly generated according to different 
values of expected selectivity, i.e., expected ratio of tuples to be returned 
by the query. For each value of expected selectivity, $10,000$ random queries 
were evaluated. The imprecision in query answering was calculated in terms 
of the median error. The results reported in Figure~\ref{fig:queryError}
show the superiority of \defense\ with respect to the other techniques;
this result is due to the use of the data quality-oriented generalization
algorithm presented in Section~\ref{subsec:bucket}.

Finally, we evaluated the number of tuples that were suppressed by
\defense\ in order to enforce the privacy requirements. Results show that
a very few number of tuples were suppressed; i.e., at most $12$ 
($< 0.25\%$) at each release.

%
%
%
%
%
%
%
%
%
%
%
%
%
%
%
%
%
%
%

 
\section{Conclusions and future work}
\label{sec:conclusion}
In this paper, we demonstrated that the correlation of sensitive values in subsequent data releases 
can be used as adversarial background knowledge 
to violate users' privacy. %
We showed that an adversary can actually obtain this knowledge by different methods.
Since serial release 
of transaction data is a common situation, the considered problem poses a very practical challenge.
We proposed a defense algorithm based on Jensen-Shannon divergence, and we showed through an extensive experimental
evaluation that other applicable solutions are not effective, while our \defense\ defense provides strong
privacy protection and good data quality, even when the adversary has more 
accurate background knowledge than the defender.

Future work includes studying the effect on privacy preservation of
compromised tuples; i.e., possibly very few tuples whose respondent is
known to the adversary.  Moreover, specific application domains (e.g.,
streaming data) often require anoymization to be performed online;
hence, a further line of investigation consists in devising protection
techniques having very low computational complexity.


\section*{Acknowledgments} 
The authors would like to thank Kristen LeFevre for providing an 
implementation of the Mondrian framework; Tiancheng Li, 
Ninghui Li and Jian Zhang for providing software modules for \bt;
and Andrea Bianchini for his extensive programming work.





\end{document}